\documentclass[twocolumn,trackchanges]{aastex701}
\usepackage{subcaption}

\usepackage[flushleft]{threeparttable}

\begin{document}

\title{Dark matter-deficient twins: FCC~224 and FCC~240 as possible analogues of NGC~1052-DF2 and DF4}

\author[orcid=0000-0003-3153-8543,gname=Bosque,sname='North America']{Maria Luísa Buzzo}
\affiliation{Astronomy Department, Yale University, 219 Prospect St, New Haven, CT 06511, USA}
\affiliation{Centre for Astrophysics and Supercomputing, Swinburne University of Technology, Hawthorn VIC 3122, Australia}
\affiliation{European Southern Observatory, Karl-Schwarzschild-Strasse 2, 85748 Garching bei M\"unchen, Germany}
\email[show]{marialuisa.buzzo@yale.edu}  

\author[orcid=0000-0002-8282-9888,gname=Bosque, sname='Sur America']{Pieter van Dokkum} 
\affiliation{Astronomy Department, Yale University, 219 Prospect St, New Haven, CT 06511, USA}
\email{fakeemail2@google.com}

\author[orcid=0000-0002-2363-5522,gname=Bosque, sname='Sur America']{Michael Hilker} 
\affiliation{European Southern Observatory, Karl-Schwarzschild-Strasse 2, 85748 Garching bei M\"unchen, Germany}
\email{fakeemail2@google.com} 

\author[orcid=0000-0001-5590-5518,gname=Bosque, sname='Sur America']{Duncan A. Forbes} 
\affiliation{Centre for Astrophysics and Supercomputing, Swinburne University of Technology, Hawthorn VIC 3122, Australia}
\email{fakeemail2@google.com} 

\author[orcid=0000-0003-2473-0369,gname=Bosque, sname='Sur America']{Aaron J. Romanowsky} 
\affiliation{Department of Physics and Astronomy, San José State University, One Washington Square, San Jose, CA 95192, USA} 
\affiliation{Department of Astronomy \& Astrophysics, University of California Santa Cruz, 1156 High Street, Santa Cruz, CA 95064, USA}
\email{fakeemail2@google.com} 

\author[orcid=0000-0003-2876-577X,gname=Bosque, sname='Sur America']{Yimeng Tang} 
\affiliation{Department of Astronomy \& Astrophysics, University of California Santa Cruz, 1156 High Street, Santa Cruz, CA 95064, USA}
\email{fakeemail2@google.com}

\begin{abstract}
The recent ``bullet-dwarf'' model proposes that high-velocity collisions between dwarf galaxies can produce stellar systems with overluminous globular clusters (GCs) and a deficiency of dark matter, as observed in the NGC~1052 group galaxies NGC~1052--DF2 and NGC~1052--DF4. We present a possible analogue system in the outskirts of the Fornax cluster: the ultra-diffuse galaxy FCC~224 and its close companion FCC~240. Using deep VLT/MUSE integral-field spectroscopy, we characterize their stellar populations, internal kinematics, and GC systems to test this formation scenario. Both galaxies exhibit low velocity dispersions. Interpreted with a standard mass estimator at the half light radius, and allowing for the known limitations associated with flattened systems, their inner dynamics are more naturally explained by stars alone than by either cuspy or cored dark matter halos. Both systems host unusually luminous GCs, closely resembling the top-heavy GCLF of the NGC~1052 pair. Moreover, FCC~224 and FCC~240 are coeval with each other, with mass-weighted stellar ages of $\sim$10~Gyr, and their GC populations share similarly old ages, in agreement with predictions of the formation scenario. Despite these similarities, FCC~224 and FCC~240 form a much tighter system than DF2 and DF4, with a projected separation of 75~kpc (compared to 240~kpc) and a relative velocity of only 16~km~s$^{-1}$ (compared to 358~km~s$^{-1}$). This distinct configuration may suggest a different present-day manifestation of the same general class of galaxies and provides additional observational constraints on models of their formation and evolution.
\end{abstract}

\keywords{\uat{Dark matter}{353}; \uat{Dwarf galaxies}{416}; \uat{Globular star clusters}{656}}

\section{Introduction} 

Ultra-diffuse galaxies (UDGs) have been the subject of intense debate over the past decade. At the heart of these discussions lie two interconnected topics: their globular cluster (GC) systems and their dark matter content. This is because UDGs are thought to span the full extremes of dark matter content in dwarf galaxies. On one end, some UDGs appear to be entirely devoid of dark matter in their inner regions \citep[e.g.,][]{vanDokkum_18, vanDokkum_19b, Danieli_19, ManceraPina_19b,ManceraPina_22, Shen_23, Buzzo_25b}; on the other, some are among the most dark-matter-dominated systems known \citep[see e.g.,][]{Gannon_20,Gannon_21,Gannon_22b,Gannon_22a,Iodice_23,Forbes_Gannon_24,Haacke_25,Buttitta_25, Mirabile_25}. Both extremes lie well outside the expected stellar mass--halo mass relation, one on the dark-matter-poor side, the other on the dark-matter-rich side, and both have been linked to the properties of their GC populations \citep{Forbes_Gannon_24}.

In the dark-matter-dominated case, the prevailing idea is that galaxies with larger numbers of GCs have correspondingly larger halo masses, following the observed GC number--halo mass relation \citep{Burkert_Forbes_20}. However, this relation is empirical and statistical, especially in the dwarf galaxy regime where many systems host only a small number of GCs. It should therefore not be interpreted as a requirement that every GC system must trace a normal dark matter halo. For example, galaxies formed from tidal debris may form clusters while containing little dark matter \citep{Fensch_19b}. In contrast, the dark-matter-deficient UDGs consistently host some GCs that are unusually luminous compared to typical populations \citep{vanDokkum_18,vanDokkum_19b,Tang_25}. Interestingly, the same process responsible for producing these overluminous GCs may also be linked to the removal or absence of dark matter in these systems \citep{Silk_19,Shin_20,Lee_21,Trujillo_Gomez_21,vanDokkum_22,Lee_24}.

Many studies have focused on the dark-matter-rich end, exploring UDG GC populations in detail. These works have shown that GC-rich UDGs tend to exhibit higher velocity dispersions, implying higher dark matter content \citep[see e.g.,][]{Iodice_23,Forbes_Gannon_24,Haacke_25,Buttitta_25}. This GC richness correlates with other UDG properties, including morphology, stellar populations, and GC formation histories \citep{Buzzo_22b,Buzzo_24,Buzzo_25a, Ferre-Mateu_23,Ferre-Mateu_25,Hartke_25}. We now have a relatively clear picture of how such galaxies can form, often interpreted as `failed galaxies' that were quenched early (although the exact quenching mechanisms are still highly debated), preventing them from forming the bulk of their expected stellar mass but leaving them with large GC numbers, and consequently, high GC mass fractions \citep[i.e., mass in GCs relative to stellar mass of the galaxy, ][]{vanDokkum_15,Danieli_22}. While the exact quenching mechanisms may depend on factors such as environment and mass, they share this same underlying principle of early truncation of star formation \citep{Danieli_22}.

The opposite extreme is far less understood. Until recently, the only known UDGs with both a top-heavy GC luminosity function (GCLF) and a lack of dark matter were NGC~1052-DF2 and NGC~1052-DF4, both associated with the NGC~1052 group. Their proximity (at least in projection) and similar properties raised the possibility that they formed together via the `bullet dwarf' scenario, where a high-speed collision between dwarf galaxies is capable of separating baryonic and dark matter while generating the pressures needed to form overluminous GCs \citep{Silk_19}. In the specific case of DF2 and DF4, this scenario was further motivated by simulations showing that high-velocity dwarf–dwarf collisions can produce multiple dark-matter–deficient fragments along a common trajectory \citep{Shin_20}. This led to the subsequent discovery of a linear trail of low-surface-brightness galaxies spanning over 2 Mpc, interpreted as fragments formed from the collisional gas stream \citep{vanDokkum_22}. However, the small sample size made it unclear whether these systems were unique to the NGC 1052 group or more widespread.

The past year has marked an important step forward in the study of these galaxies. A dark matter--deficient dwarf galaxy, FCC~224, has now been identified outside the NGC~1052 group \citep{Tang_25,Buzzo_25b}. At the same time, idealized simulations have begun to reproduce systems with similar internal properties \citep{Lee_24}, and new observational studies have started to connect these galaxies to specific stellar populations and morphologies \citep{Tang_24}. We are therefore now in a position to compare individual systems and assess whether they represent additional outcomes of the bullet-dwarf mechanism or arise from other processes entirely, such as tidal stripping \citep{Ogiya_18,Moreno_22}, tidal heating \citep{Carleton_21}, etc.

In this work, we extend the study of FCC~224 using deep VLT/MUSE observations of both the galaxy itself and a projected close companion in the Fornax cluster, FCC~240. Together, these two galaxies form a pair that may represent a possible analogue of DF2 and DF4 in the NGC~1052 group. FCC~224 has previously been studied in detail using \textit{HST} imaging by \cite{Tang_25} and Keck/KCWI spectroscopy by \cite{Buzzo_25b}. The photometric analysis revealed a population of $12 \pm 1$ GC candidates and a top-heavy GCLF, while the spectroscopic observations showed that both the stellar body and the GCs have very low velocity dispersions consistent with little or no dark matter. That study relied on six spectroscopically confirmed GCs and, because of the small number of tracers, could only place upper limits on the velocity dispersion.

The MUSE observations presented here probe a larger number of GC candidates and extend to larger radii in the stellar body, providing improved statistics and an independent dataset with which to measure the internal kinematics. By combining the new MUSE data for FCC~224 with similar observations of FCC~240, we carry out a homogeneous analysis of their stellar populations, GC systems, and structural properties, and place the FCC~224/FCC~240 system in the broader context of dark matter--deficient dwarf galaxies.

\begin{figure*}
    \centering
    \includegraphics[width=\textwidth]{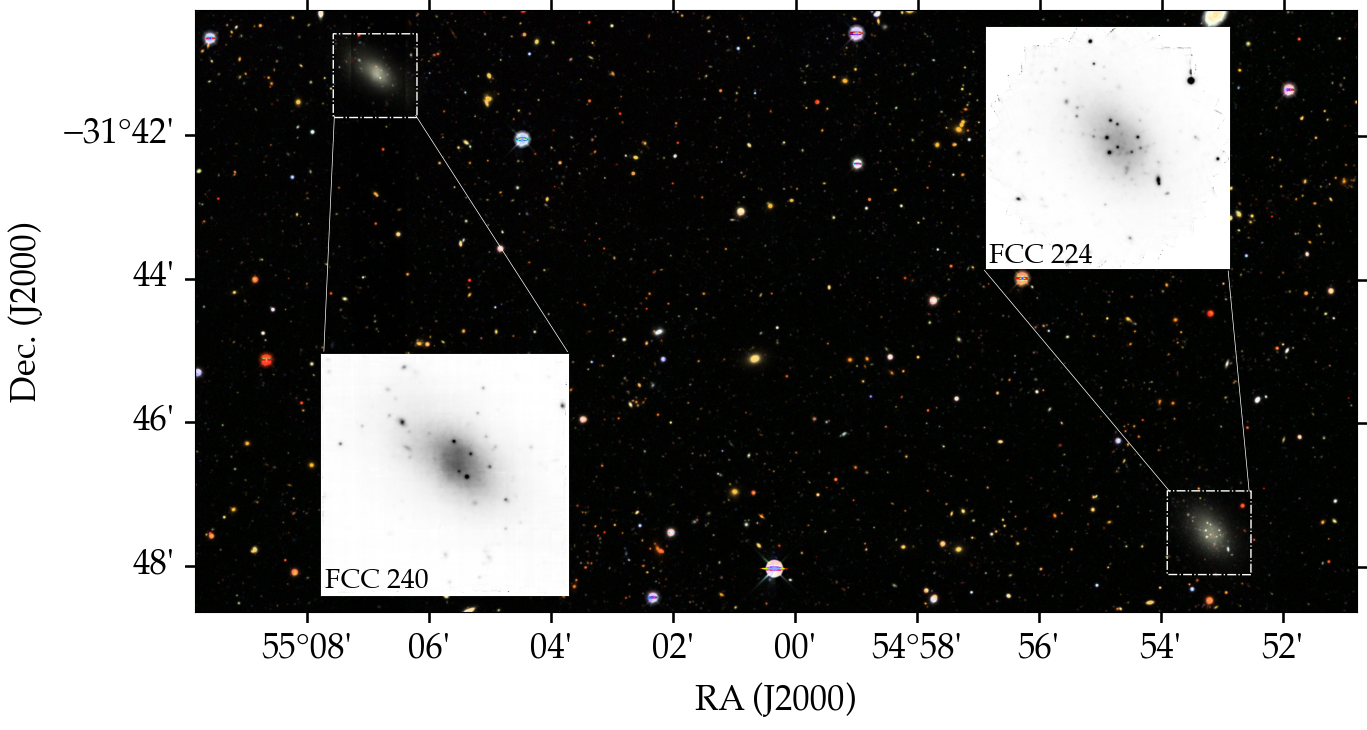}
    \caption{
    Local environment and MUSE coverage of the FCC~224/FCC~240 system.
    This composite image shows our MUSE pointings (white rectangles) overlaid on a deep DECaLS DR10 sky survey image, revealing the vicinity of the two dwarf galaxies in the outskirts of the Fornax cluster. The MUSE fields cover the galaxies out to more than 2 effective radii $(R_{\rm e})$. 
    %The alignment of their major axes and similar elongation, combined with their close projected separation, suggest a physical connection and potential common origin. 
    }
    \label{fig:MUSE_Legacy}
\end{figure*}

\section{Data and Analysis} 
\label{sec:data}

In this section, we describe the MUSE data collection and reduction for both galaxies, as well as the sky subtraction strategy adopted. The full spectral fitting method applied to the datacubes used to extract kinematics and stellar population information is also detailed.

\subsection{Observations and Data Reduction}

Integral field spectroscopy of FCC~224 and FCC~240 was obtained with VLT/MUSE under ESO programme ID 114.27CM.001 (PI: Buzzo). Observations for FCC~224 were obtained in dark time from 2024 October 7 to 2025 January 4, and for FCC~240 in dark time from 2025 January 21 to 2025 January 27. For FCC~224, a total of 13 hours of on-target exposure time were divided into observing blocks (OBs) of one hour, each comprising four dithered sub-exposures. Each observing block was tilted by 20 degrees from the previous one to average out slicer patterns when combining the deep cube. For FCC~240, there were four hours of on-target exposure time divided into four OBs, each comprising four dithered sub-exposures. The observing blocks were tilted by 90 degrees again to average out slicer patterns.

The observations were conducted without adaptive optics in Wide Field Mode, yielding a $1 \times 1$ arcmin$^2$ field-of-view (FoV) with $0.2 \times 0.2$ arcsec$^2$ spatial sampling. The wavelength coverage spans 4800 to 9300 \AA, with a spectral resolution ranging from 69 km s$^{-1}$ (at 5000 \AA) to 46 km s$^{-1}$ (at 7000 \AA) \citep{Bacon_17,Guerou_17,Emsellem_19}. 

The data were reduced using the MUSE pipeline \citep[v2.14.10,][]{Weilbacher_16} within the \texttt{ESOREFLEX} environment \citep{Freudling_13}. The standard reduction steps included bias and flat-field corrections, line spread function (LSF) creation, wavelength calibration, and illumination correction.

For FCC~224, two different reductions were applied. We first combined all of the 52 exposures to get the maximum information for the stellar body of the galaxy. To study the GCs, we selected only exposures with seeing $\lesssim$1.0 arcsec, which reduced the number of exposures to 30. For FCC~240, a single reduction was applied to all exposures, with a final seeing of $\sim$1.0 arcsec. 

Sky subtraction was optimized following the method described in \cite{Iodice_23}, \cite{Buttitta_25}, \cite{Hartke_25}, and \cite{Buzzo_25c}. After an initial reduction without any sky subtraction, a custom sky mask was generated using \texttt{SExtractor} \citep{Bertin_02}, which excluded spaxels associated with the galaxy, GCs, and foreground/background objects. This mask was then applied in subsequent reduction steps, with a sky fraction of 75\% for estimating the background. Finally, residuals were cleaned with \texttt{ZAP} \citep{Soto_16} to produce the final datacubes. The MUSE white-light images of both FCC~224 and FCC~240 are shown in Figure~\ref{fig:MUSE_Legacy}.

\begin{figure*}[t!]
    \centering
    \includegraphics[width=\textwidth]{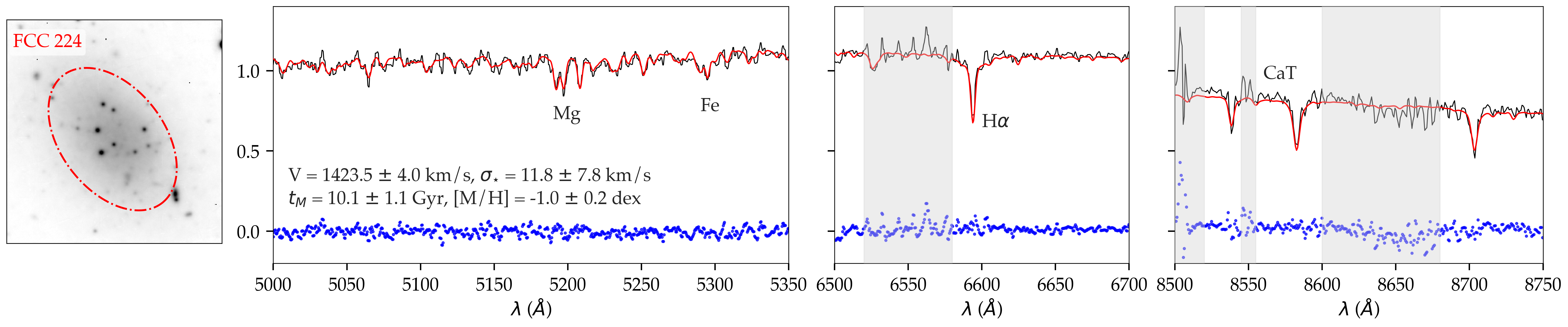}
    \includegraphics[width=\textwidth]{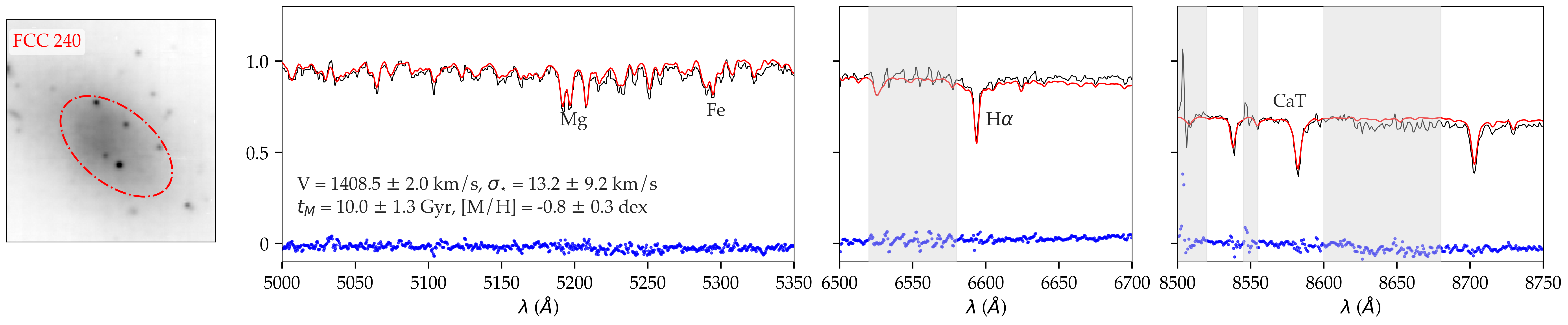}
    \caption{Integrated stellar spectra of FCC~224 (top) and FCC~240 (bottom).
    Left panels show the MUSE white-light images with the $1\,R_{\rm e}$ elliptical apertures used to extract the integrated spectra marked in red.
    The right panels show representative spectral regions, with the observed spectra in black, the best-fitting models in red, and the residuals in blue.
    Grey bands indicate spectral regions that were masked during the fitting.}
    \label{fig:kinematics_stars}
\end{figure*}

\subsection{Full spectral fitting of stellar body and GCs}

Our analysis of the stellar body of both galaxies was performed using the Data Analysis Pipeline \citep[\texttt{DAP};][]{Westfall_19, Belfiore_19}, which serves as a wrapper for the \texttt{pPXF} stellar kinematics fitting routine \citep{Cappellari_04, Cappellari_17}. After generating a segmentation map with \texttt{SExtractor} to mask all sources (including GC candidates) in the foreground and background, we extracted the integrated spectrum of the stellar body of each galaxy within one $R_{\rm e}$ using an ellipse following the properties obtained for FCC~224 by \citet{Tang_25}, and for FCC~240 using the values obtained in Appendix \ref{sec:appendix_fcc240}. The final S/N of the spectra, measured with the \texttt{DER\_SNR} algorithm, are 53 \AA$^{-1}$ and 84 \AA$^{-1}$ for FCC~224 and FCC~240, respectively. The stellar continuum was modelled using the MILES simple stellar population templates \citep{Vazdekis_15}, which has a resolution similar to that of MUSE ($\sim$ 2.5 \AA) and provided robust measurements of the line-of-sight velocity, and stellar population parameters.

Recovering the velocity dispersion of these galaxies with MUSE is challenging due to its instrumental resolution of $\sim$50 km s$^{-1}$, which limits the reliability of measurements below this threshold. However, previous works by \cite{Emsellem_19}, \cite{Buttitta_25}, and \cite{Iodice_23} have demonstrated that a careful characterization of the LSF can constrain the velocity dispersion well below the resolution limit. Simulations in \cite{Iodice_23}, for example, show that high signal-to-noise MUSE data (S/N$>$20 \AA$^{-1}$) can typically recover input velocity dispersions in the range of 10--20 km s$^{-1}$. In our case, to ensure reliable measurements and realistic uncertainty estimates, we applied small perturbations to the recovered LSF and performed 1000 Monte Carlo iterations with \texttt{pPXF}. This approach allowed us to recover velocity dispersions with realistic, albeit large, uncertainties that properly reflect the limitations of the data.

The results of this analysis are shown in Figure~\ref{fig:kinematics_stars}. For FCC~224, we fitted the integrated spectra within 1~$R_{\rm e}$ and recovered a systemic radial velocity of $V_{\rm sys} = 1423.5 \pm 4.0$ km s$^{-1}$, a stellar velocity dispersion of $\sigma_\star = 11.8 \pm 7.8$ km s$^{-1}$, a mass-weighted age of $10.1 \pm 1.1$ Gyr, and a metallicity of [M/H] $= -1.0 \pm 0.2$ dex. The stellar kinematic and population properties derived here can be directly compared with previous studies of FCC~224. Using Keck/KCWI spectroscopy, \citet{Buzzo_25b} measured a systemic velocity of $V_{\rm sys} \simeq 1405$ km s$^{-1}$, a stellar velocity dispersion of $\sigma_\star = 8.5^{+2.3}_{-3.1}$ km s$^{-1}$, and found a stellar age of $\sim$10 Gyr, and a metallicity of ${\rm [M/H]}\sim-1.0$, consistent with a single early formation epoch. Using a combination of photometric and spectroscopic data, \citet{Tang_25} obtained $V_{\rm sys} = 1422 \pm 13$ km s$^{-1}$, and old and metal-poor stellar populations, in excellent agreement with the MUSE results presented here. The modest offset in systemic velocity between the KCWI results from \cite{Buzzo_25b} and MUSE measurements is likely driven by small systematic effects, potentially related to wavelength calibration. In addition, the KCWI observations targeted only the central regions of FCC~224 with the medium slicer, while both the MUSE data and the analysis of \citet{Tang_25} probe a larger area. 

For FCC~240, we measure within 1~$R_{\rm e}$ a radial velocity of $V_{\rm sys} = 1408.5 \pm 5$ km s$^{-1}$, a velocity dispersion of $\sigma_\star = 13.2 \pm 9.2$ km s$^{-1}$, an age of $10.0 \pm 1.3$ Gyr, and [M/H] $= -0.8 \pm 0.3$ dex, revealing a close proximity to FCC~224 in velocity space, and similar stellar populations. 

After analyzing the integrated spectra, we looked into the internal structure of the galaxies. For this, we Voronoi-binned the data out to $2~R_{\rm e}$, requiring a minimum S/N of 30 \AA$^{-1}$ per bin. This analysis revealed no signs of rotation in either galaxy. We also computed the projected angular momentum parameter $\lambda_{R_{\rm e}}$ following \citet{Emsellem_11}. Because the velocity dispersions of individual Voronoi bins are not reliable at the very low values measured here, we use the integrated velocity dispersion measured within $1R_{\rm e}$ for each galaxy, rather than the bin by bin dispersions. This gives $\lambda_{R_{\rm e}}<0.1$ for FCC~224 and FCC~240, consistent with their lack of measurable rotation. Applying the same procedure to archival MUSE data for DF2 and DF4 gives similarly low values. The full calculation and its limitations are described in Appendix~\ref{sec:lambda_appendix}. Both age and metallicity gradients were found to be flat out to $2~R_{\rm e}$. These properties are unusual for dwarf galaxies, which often show either rotational support or clear population gradients 
\citep[see e.g.,][]{Lipka_24,Buttitta_25,Ferre-Mateu_25}. The absence of rotation and gradients is consistent with a rapid formation event that efficiently mixed the stellar populations, rather than forming stars in a disk.

\begin{figure*}[t!]
    \centering
    \includegraphics[width=\textwidth]{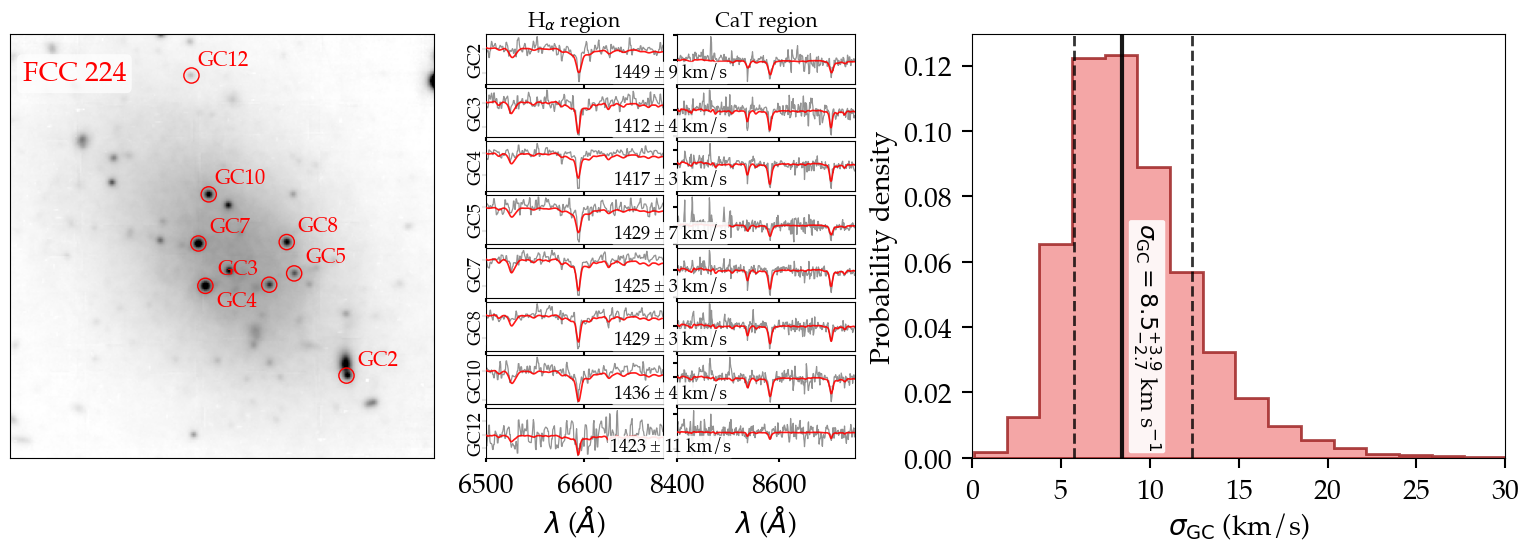} 
    \includegraphics[width=\textwidth]{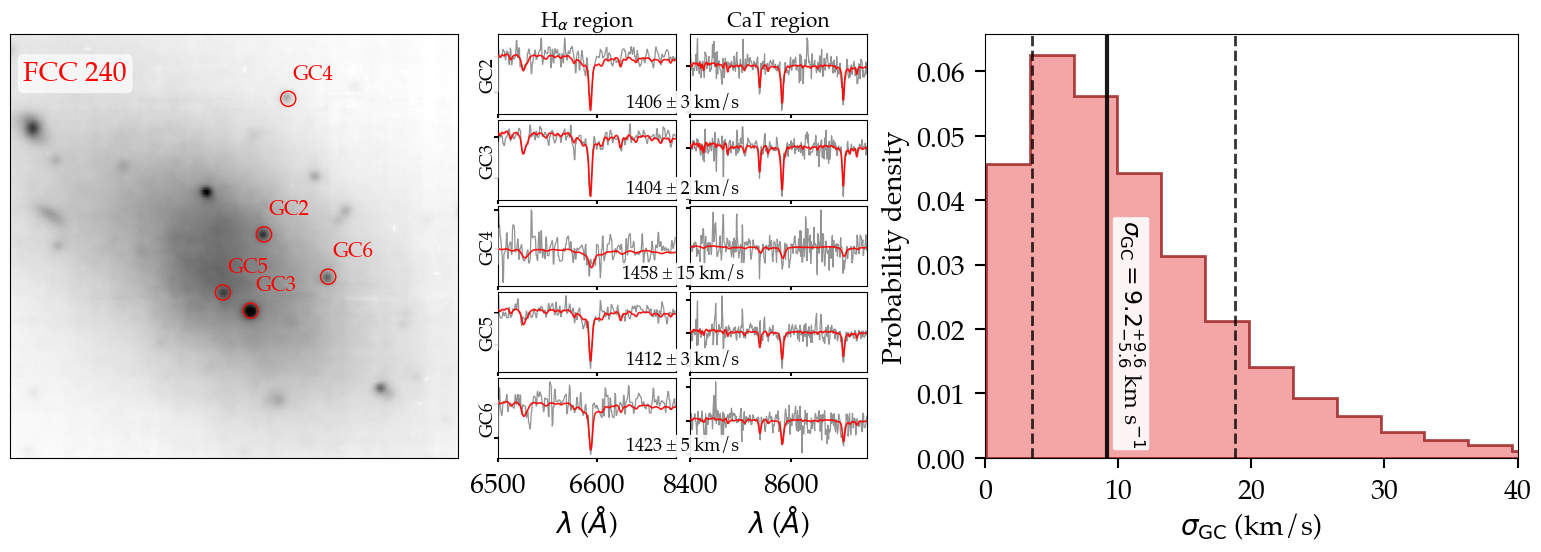} 
    \caption{GC kinematics and dynamics in FCC~224 (top) and FCC~240 (bottom). The left panels show the white-light image of the galaxies with the confirmed GCs marked in red. The middle panels show individual GC spectra along with their recovered velocities. The right panels show the MCMC analysis used to estimate the velocity dispersion of the GC systems.}
    \label{fig:kinematics_GCs}
\end{figure*}

We next analyzed the GC systems to confirm membership and to derive their kinematics and stellar population properties. A source was considered a confirmed GC if its radial velocity lay within 100 km s$^{-1}$ of the host galaxy systemic velocity -- although in practice, all confirmed clusters lie much closer, typically within $\sim$20 km s$^{-1}$ of the host.

For FCC~224, we used the datacube constructed from the best-seeing exposures and examined the GC candidates identified from deep \textit{HST} imaging by \citet{Tang_25}, who statistically estimated that the galaxy should host $12 \pm 1$ GCs. Of these, nine candidates have spectra with sufficient quality (S/N$>5$ \AA$^{-1}$) to allow a reliable velocity measurement; GC candidates numbered 1, 6, and 9 in the nomenclature of \citet{Tang_25} have S/N below this threshold and were therefore not tested. Eight of the nine tested candidates were confirmed as members. The ninth candidate, GCC11, has a measured velocity of $1309.3 \pm 125.4$ km s$^{-1}$, which is formally consistent with the systemic velocity of the galaxy within the uncertainties, but was excluded because of its large velocity uncertainty and because it does not strictly satisfy the adopted membership criterion of $\pm100$ km s$^{-1}$. We therefore conservatively adopt eight confirmed GCs for FCC~224. Notably, all six GCs confirmed with KCWI in \cite{Buzzo_25b} around FCC~224 were also confirmed with MUSE.

For FCC~240, GC candidates were pre-selected using DECaLS DR10 imaging based on magnitude, color, and morphological criteria. Six sources meet the GC selection requirements, and all have spectra with S/N$>5$ \AA$^{-1}$. Five of the six were confirmed members of FCC~240, with the sixth having a redder color and being found to be $\sim200$ km s$^{-1}$ in the background of FCC~240, thus it is likely a background galaxy. The full photometric selection and estimate of the total GC number for FCC~240 are described in Appendix~\ref{sec:appendix_fcc240}. The confirmed GCs for both galaxies are shown in Figure~\ref{fig:kinematics_GCs}, together with their measured radial velocities, also listed in Tables~\ref{tab:FCC240} and \ref{tab:FCC224}.

To estimate the velocity dispersion of the GC systems, we employed the same technique used in \cite{Toloba_23}, \cite{Haacke_25}, and \cite{Buzzo_25b}, applying Markov Chain Monte Carlo (MCMC) with Jeffreys priors. We fixed the systemic velocity to that of the host galaxy and sampled only the velocity dispersion. The results, shown in Figure~\ref{fig:kinematics_GCs}, yield velocity dispersions of $\sigma_{\mathrm{GC}} = 8.5^{+3.9}_{-2.7}$ km s$^{-1}$ for the eight confirmed GCs in FCC~224 and $\sigma_{\mathrm{GC}} = 9.2^{+9.6}_{-5.6}$ km s$^{-1}$ for the five GCs in FCC~240. The agreement between the stellar and GC velocity dispersions for each galaxy reinforces the reliability of the stellar measurement, even when probed below the instrumental resolution of MUSE. 

A caveat to the GC-based dispersions is that the spectroscopic samples are dominated by the brightest clusters. In FCC~224, previous work has shown evidence for mass segregation, with the most massive GCs being more centrally concentrated \citep{Tang_25}. The confirmed GCs used here have a half-number radius of $\sim0.5\,R_{\rm e}$ relative to the stellar light. If the GC system is radially segregated, the velocity dispersion measured from the bright GCs is expected to be lower than that of the full GC population, and potentially lower than the stellar dispersion, even in the absence of dark matter. For this reason, the GC dispersions should not be interpreted in isolation as direct probes of the total gravitational potential. Instead, the key result is the similarity between the stellar and GC kinematics, and the fact that both remain well below the dispersions expected for typical dwarf galaxies of similar stellar mass.

We then stacked the eight confirmed GCs in FCC~224 and the five in FCC~240 to derive their ages and metallicities, finding that both are consistent within the uncertainties with a mass weighted age of $t_{\rm M}=10.1\pm1.2$~Gyr and a metallicity of ${\rm [M/H]}=-1.0\pm0.2$~dex, equivalent to those of the diffuse stellar bodies discussed above.

Given the different spectral resolutions and wavelength coverage of MUSE and KCWI, we do not combine the GC spectra or results obtained by \citet{Buzzo_25b} for FCC~224 with those presented here. Nevertheless, the inferred GC ages and metallicities agree within the uncertainties between the two studies.

\section{Dark Matter Deficiency}
\label{sec:darkmatter}

The very low line-of-sight velocity dispersions measured for both the stellar bodies and the GC systems of FCC~224 and FCC~240 immediately suggest that their inner gravitational potentials are unusually shallow for their stellar masses. However, low velocity dispersions alone do not automatically imply a lack of dark matter: they must be interpreted in comparison with theoretical expectations for galaxies of similar stellar mass embedded in dark matter halos. In this section, we therefore test whether the observed kinematics are consistent with three broad classes of models: galaxies embedded in standard cuspy dark matter halos, galaxies hosted by cored dark matter halos, and systems in which the gravitational potential within the stellar body is dominated by the stars, with little or no dark matter contribution.

\subsection{Expected velocity dispersions from different mass models}

To place the measured dispersions in context, we compute the expected line-of-sight velocity dispersions at one effective radius. This is the radius where mass estimates for dispersion supported systems are least sensitive to the usual mass anisotropy degeneracy \citep{Wolf_10}. This does not remove all possible effects of flattening or triaxiality, but it makes the comparison less sensitive to the unknown orbital structure than estimates at smaller or larger radii. For the cuspy-halo case, we assume that each galaxy resides in a dark matter halo whose mass follows the stellar mass--halo mass relation of \citet{Moster_13}. The halo density profile is taken to be NFW-like \citep{NFW_97}, with halo concentrations drawn from the mass--concentration relation of \citet{Dutton_Maccio_14}, including its intrinsic scatter. The stellar mass distribution is described by a Sérsic profile, with structural parameters and stellar masses taken from \citet{Tang_25} for FCC~224 and from Appendix~\ref{sec:appendix_fcc240} for FCC~240. In both cases, the galaxies are modeled with \texttt{GALFIT}, and stellar masses are derived from spectral energy distribution fitting using \texttt{Prospector} \citep{Johnson_21}, assuming an exponentially declining star formation history and a \citet{Kroupa_01} initial mass function. 

We also consider cored dark matter halos, which have reduced central dark matter densities relative to cuspy profiles. These models are constructed by modifying the inner density profile of the halo while keeping the total halo mass fixed to the value implied by the stellar mass--halo mass relation. We adopt a core radius of $r_c = 3R_e$, toward the upper end of values typically found for dwarf galaxies, in order to test whether large cores proposed in the literature as a possible explanation for low velocity dispersions can reconcile the observations. 
% As expected, introducing such a core lowers the predicted velocity dispersion at the effective radius by reducing the dark matter contribution to the gravitational potential in the inner regions.

Finally, for the no--dark--matter case, we compute the expected velocity dispersion assuming that the gravitational potential within the stellar body is dominated by the stellar mass alone. The stellar mass distribution is described by a Sérsic profile with the observed structural parameters of each galaxy, and the predicted dispersion reflects the self-gravity of the stars. This model does not imply that the galaxy contains no dark matter at all, but rather that any dark matter present contributes negligibly to the dynamics within the probed region.

\begin{figure}
\centering
\includegraphics[width=\linewidth]{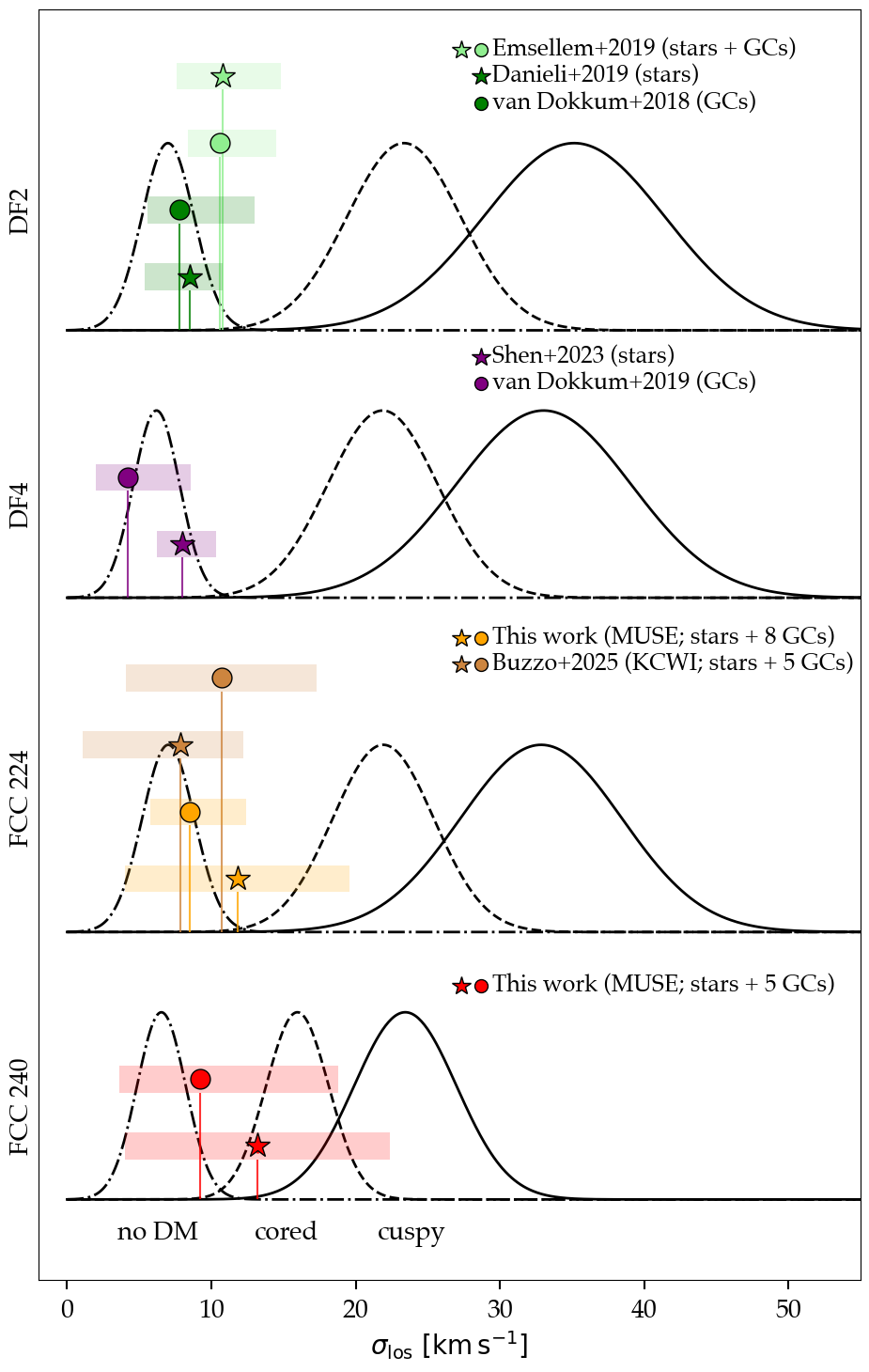}
\caption{
Velocity dispersion constraints for FCC~224 and FCC~240 compared to theoretical expectations. The plot shows the measured velocity dispersions from stars (star symbols) and GCs (circles) compared to the expected ranges for systems with cuspy NFW dark matter halos (solid Gaussian), cored dark matter halos assuming a core radius of $r_c = 3~R_{\rm e}$ (dashed Gaussian), and for systems without dark matter (dot--dashed Gaussian). Vertical lines are included to guide the eye. Error bars indicate $1\sigma$ uncertainties. The measurements for FCC~224 lie well below the expectations for normal dark matter halos, even when cored models are considered, and closely follow the no-dark-matter predictions, similar to NGC~1052-DF2 and DF4 (also shown). For FCC~240, the stellar dispersion is formally consistent with the cored-halo expectations, although the large uncertainties and compactness of the galaxy make this constraint weaker.}
\label{fig:dm_mass}
\end{figure}

Figure~\ref{fig:dm_mass} compares the observed stellar and GC velocity dispersions for FCC~224, FCC~240, DF2, and DF4 to the expectations from these three models. Rather than showing broad shaded bands that span the full range of possible halo parameters, as is commonly done in the literature \citep{Danieli_19,Shen_23,Buzzo_25b}, we represent each model as a Gaussian curve representing the expected distribution of velocity dispersions for a given model, including the intrinsic scatter arising from the assumed halo relations.

% This way of visualizing the comparison makes the degree of dark matter deficiency more transparent. Shaded bands can give the impression that models overlap with the data simply because they cover a wide parameter space, even when the bulk of the expected values lie far from the measurements. By contrast, the Gaussian representation highlights where most of the model probability density lies. In particular, it becomes immediately clear that the measured dispersions for FCC~224 lie well below the expectations for normal cuspy halos and remain low even relative to the cored-halo expectations, as is the case for DF2 and DF4. FCC~240 is a more ambiguous case.

\subsection{Are cored halos sufficient?}

Recent hydrodynamical simulations predict that galaxies at the stellar masses of UDGs may host dark matter halos with central cores rather than cuspy density profiles \citep{Tollet_16}. This expectation has been tested observationally for samples of UDGs and found to be broadly consistent with the data \citep{Forbes_Gannon_24}. Considering cored halos for FCC~224 and FCC~240 is therefore well motivated given their stellar masses.

In principle, introducing a core into the dark matter halo can lower the predicted velocity dispersion at the effective radius. However, reproducing dispersions as low as those observed here while maintaining a halo mass consistent with the stellar mass--halo mass relation generally requires cores that are very large compared to the size of the stellar body. In Figure~\ref{fig:dm_mass} we show cored-halo predictions adopting a core radius of $r_c = 3R_e$, already at the upper end of values commonly assumed for dwarf galaxies \citep[e.g.,][]{Read_16,Tollet_16,DiCintio_17}. Even with this choice, the predicted velocity dispersions remain significantly higher than those observed for FCC~224, similar to DF2 and DF4. Reconciling the observed dispersions would require core radii several times larger ($\gtrsim 5$--$10\,R_e$), implying extremely low central dark matter densities and very extended halos relative to the stellar body.

This difficulty has been emphasized repeatedly in the literature. For example, comparisons with Local Group dwarf galaxies of similar stellar mass done by \cite{vanDokkum_18} show that even quiescent systems typically exhibit stellar velocity dispersions of $\sim20-30$ km s$^{-1}$ and large dynamical-to-stellar mass ratios within their half-light radii \citep[e.g.,][]{Walker_09, Battaglia_13, Simon_19, Read_19}. While it is possible to construct cored-halo models that formally match the observed dispersions, doing so requires pushing the halo structure into an extreme regime that is not commonly inferred for dwarf galaxies. In addition, the presence of a large dark matter core would be in tension with the dynamical friction signatures inferred for the GC system of FCC~224 \citep{Tang_25}, as well as with the relatively central concentration of its GCs. In cored halos, GCs are expected to stall at approximately the core radius during orbital decay, as observed in the Fornax dwarf spheroidal galaxy \citep[e.g.,][and references therein]{Amorisco_14}. The apparent absence of such stalling in FCC~224 thus disfavors models with large cores.

More flexible dynamical modeling reaches a similar conclusion. For example, \citet{Aditya_24} showed that cored halos with halo masses consistent with the stellar mass--halo mass relation can fit the DF2 data only if the dark matter distribution is extraordinarily extended, such that the stellar mass dominates the gravitational potential throughout the observed region. In these cases, the kinematics remain equally well explained by models with little or no dark matter within the stellar body. Similarly, \citet{Wasserman_18} found that matching the observed velocity dispersion requires cores that are sufficiently large to substantially reduce the dark matter contribution in the inner regions, again leaving the stellar component to dominate the dynamics within the half-light radius.

\subsection{Interpretation for FCC~224 and FCC~240}

For FCC~224, both the stellar and GC velocity dispersions lie far below the expectations for cuspy halos and remain low even relative to the cored predictions. They are consistent with the stellar mass dominating the gravitational potential within $R_{\rm e}$, closely resembling the behavior observed for DF2 and DF4 \citep{vanDokkum_18, Danieli_19, Shen_23}.

For FCC~240, the situation is less clear because the galaxy is more compact and the uncertainties on the velocity dispersion are larger. While the stellar dispersion is formally consistent with the cored-halo expectations, it is also consistent with the no-dark-matter case. Importantly, even under the most conservative interpretation, FCC~240 is clearly inconsistent with the expectations for a typical cuspy halo at its stellar mass.

In all four galaxies, the measured velocity dispersions are slightly above the no-dark-matter predictions, but still consistent with a dark-matter-deficient scenario. We are not claiming the absence of dark matter, only that these systems contain less than typical dwarfs of similar stellar mass within the inner regions probed.

\subsection{Dynamical mass estimates}

To quantify the degree of dark matter deficiency implied by the low velocity dispersions, we convert the measurements into dynamical mass estimates within one effective radius using the estimator of \citet{Wolf_10}. This estimator was derived for spherical, dispersion-supported galaxies, but has been shown to hold for flattened galaxies when evaluated at the circularized $R_{\rm e}$, provided the stellar and dark matter distributions share a similar flattening \citep{Sanders_16,GonzalezSamaniego_17}.

For FCC~224, the stellar velocity dispersion $\sigma_\star = 11.8 \pm 7.8$~km~s$^{-1}$ and effective radius $R_{\rm e} = 1.89$~kpc yield a dynamical mass within $r_{1/2}$ of
$M_{\rm dyn}(<r_{1/2}) = 1.1^{+1.9}_{-1.0} \times 10^8\,M_\odot$.
With a stellar mass of $M_\star = 1.7 \times 10^8\,M_\odot$, this corresponds to $(M_{\rm dyn}/M_\star)_\star = 1.3^{+2.3}_{-1.1}$.
Using the GC dispersion yields $(M_{\rm dyn}/M_\star)_{\rm GC} = 0.66^{+0.75}_{-0.35}$.

For FCC~240, the stellar dispersion $\sigma_\star = 13.2 \pm 9.2$~km~s$^{-1}$ and effective radius $R_e = 1.0$~kpc imply
$M_{\rm dyn}(<r_{1/2}) = 7.2^{+13.5}_{-6.5} \times 10^7\,M_\odot$,
corresponding to $(M_{\rm dyn}/M_\star)_\star = 1.78^{+3.34}_{-1.61}$.
Using the GC velocity dispersion yields a consistent value of
$(M_{\rm dyn}/M_\star)_{\rm GC} = 1.2^{+1.4}_{-0.7}$.
While the constraints for FCC~240 are necessarily weaker because of its compact size and larger uncertainties, both tracers remain consistent with a low dark matter content within one effective radius.

% Within the uncertainties, the dynamical masses of both FCC~224 and FCC~240 are consistent with their stellar masses alone, with no strong evidence for a significant dark matter contribution inside the half-light radius. 
The derived dynamical-to-stellar mass ratios for FCC~224 and FCC~240 are far lower than those observed in most dwarf galaxies, which typically have $M_{\rm dyn}/M_\star \sim 5$--50 within their half-light radii \citep{Walker_09, Battaglia_13, Simon_19, Read_19,Pace_25}. In this sense, the key result is not the precise inner slope of the dark matter density profile, but the absence of the strong dark matter signature that is otherwise ubiquitous in dwarf galaxies at these stellar masses. 

As noted above, the \citet{Wolf_10} estimator was derived for spherical, dispersion-supported systems and assumes a roughly flat velocity dispersion profile near the half-light radius. The observed flattening of FCC~224 and FCC~240 should therefore be taken into account when interpreting the recovered dynamical masses. We emphasize, nevertheless, that studies of dispersion-supported dwarf galaxies show that the \cite{Wolf_10} estimator remains robust for flattened systems when evaluated at the circularized $R_{\rm e}$ \citep{Sanders_16,GonzalezSamaniego_17}. These works find that flattening and projection effects introduce uncertainties of order tens of percent, which are small compared to the discrepancy between the observed velocity dispersions and those expected for normal dark matter halos.

The GC populations of these galaxies provide a different way to view the problem. Using the empirical relation of \citet{Burkert_Forbes_20}, the total GC numbers of FCC~224 ($N_{\rm GC}=12\pm1$) and FCC~240 ($N_{\rm GC}=8\pm3$) would suggest halo masses of $\sim10^{10}-10^{11}\,M_\odot$. Such halos would normally imply much larger velocity dispersions than those measured here. This apparent tension, however, does not necessarily contradict a dark-matter-deficient interpretation. The GC number--halo mass relation is empirical and, in the dwarf regime, is expected to hold only statistically. It is also not a physical requirement that GCs form exclusively in galaxies that retain normal dark matter halos \citep{Fensch_19b}. We therefore interpret the GC richness of FCC~224 and FCC~240 as further evidence that these systems are unusual, rather than as evidence against their low inner dark matter content.

\section{An Overluminous Globular Cluster Population}
\label{sec:gcpopulations}

\begin{figure}
\centering
\includegraphics[width=\columnwidth]{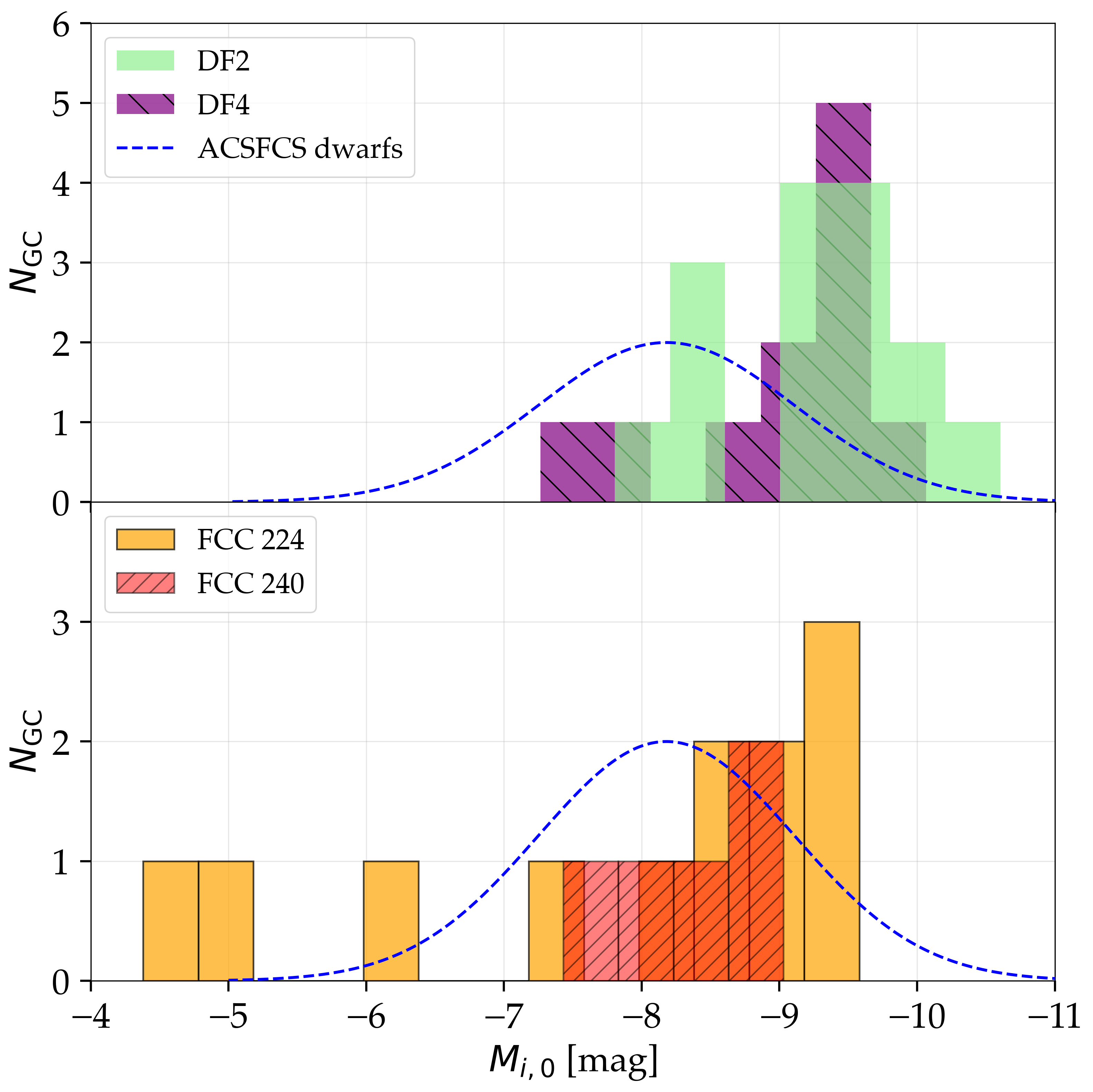}
\caption{
GCLF of the dark matter deficient dwarfs compared with the canonical GCLF of dwarf galaxies. 
Top: distributions of $i$ band magnitudes for the GCs in DF2 (green) and DF4 (purple). The blue dashed curve shows the standard GCLF for dwarf galaxies in the Fornax cluster \citep{Jordan_15}. 
Bottom: GCLFs of FCC~224 (orange) and FCC~240 (red) including all GC candidates, not only the confirmed ones. Both show a clear shift toward brighter magnitudes and a dearth of faint clusters. 
}
\label{fig:gclf}
\end{figure}

Having quantified the dark matter deficiency, we now return to the GC systems of the two galaxies. A defining property of the GCs around FCC~224, as shown by \citet{Tang_25}, and of those around DF2 and DF4 \citep{Shen_21a}, is that they are unusually bright. Their GCLF is shifted toward brighter magnitudes compared to the canonical GCLF, which peaks at $M_i \approx -8.2$ mag, as measured for Fornax dwarf galaxies by \citet{Jordan_15}. We note that \citet{Jordan_15} used the F850LP band from the ACS Fornax Cluster Survey, with the conversion to the $i$ band adopted here taken from \citet{Tang_25}.

Here we extend this analysis to the GCs around FCC~240 to test whether they exhibit the same behavior. Figure~\ref{fig:gclf} shows the GCLFs of all four galaxies. The GC magnitudes for DF2, DF4 \citep{vanDokkum_22b}, and FCC~224 \citep{Tang_25} measured in the F814W band are treated here as equivalent to the $i$ band for comparison with FCC~240, which does not have \emph{HST} imaging.

The GC system of FCC~240 closely resembles that of FCC~224. Its clusters have a median absolute magnitude of $M_{i,0} \approx -8.5$ mag, assuming a distance of 20 Mpc for the galaxy, nearly one magnitude brighter than the peak of the standard GCLF. This value closely matches those measured for DF2 and DF4 \citep{vanDokkum_18,vanDokkum_19b,vanDokkum_22b}, indicating that FCC~240 also hosts an abundance of overluminous GCs.

The apparent paucity of faint or ``normal'' GCs should, however, be interpreted with caution. Deep photometric studies of DF2 and DF4 have identified additional faint GC candidates \citep{Shen_21a, vanDokkum_22b}, and a similar population has been reported for FCC~224 \citep{Tang_25}. For FCC~240, the limited depth of the DECaLS DR10 imaging restricts our sensitivity to the faint end of the GCLF. An exploration of pseudo-$g$, $r$, and $i$ MUSE bands showed the same limitation, with a depth comparable to DECaLS DR10 and the limited FoV making background and completeness estimates more uncertain. Ongoing and upcoming \emph{Euclid} observations of the Fornax cluster are beginning to characterize GC systems around dwarf galaxies and will soon reach the outskirts of the cluster, where FCC~224 and FCC~240 reside, providing more complete GC catalogues \citep{Saifollahi_25}. While deeper data will be required to fully assess the faint GC population, the existing observations already hint at an excess of luminous GCs relative to typical dwarf galaxies.

A top-heavy GCLF is one of the key predictions of the bullet-dwarf scenario. The presence of such bright GCs in both galaxies therefore supports the idea that they formed through this type of event. The close similarity between their GC properties and those of DF2 and DF4 further strengthens this interpretation.

\section{A Coeval Formation for Stars and GCs}
\label{sec:compare}

One of the strongest predictions of the bullet-dwarf scenario is that the stars and GCs form at the same time. A high-velocity collision is expected to trigger a short and intense burst of star formation that produces both the diffuse stellar body and unusually massive clusters \citep{Lee_21}. Observational evidence for this behavior has already been found in DF2, DF4, and FCC~224, using both photometry and spectroscopy. For example, \citet{Tang_25} and \citet{vanDokkum_22b} showed that the GCs around FCC~224 and DF2/DF4, respectively, are remarkably monochromatic, exhibiting a very narrow color range compared to GC systems in other dwarf galaxies in the Virgo and Fornax clusters. DF2 and DF4 were also shown to host similar stellar populations, both internally and relative to their GC systems. Beyond this, \citet{Tang_24} demonstrated that the stellar populations of most galaxies along the proposed trail in the NGC~1052 group are broadly consistent with one another, and \citet{Buzzo_23} found that the galaxies along the trail that host GCs all contain GCs with similar colors.

In this work, we test whether FCC~240 exhibits the same behavior, and we compare the diffuse stellar light and GC populations across all four galaxies. We focus on colors rather than derived stellar population parameters for several reasons: (1) colors are direct observables; (2) different methods, instruments, and modelling assumptions can lead to systematic differences in inferred ages and metallicities; and (3) colors allow a straightforward comparison of the GC color dispersion. This comparison is shown in Figure~\ref{fig:stars_gcs_comparison}. The $g-i$ colors for FCC~240 are measured in Appendix~\ref{sec:appendix_fcc240}. For DF2 and DF4, the colors were measured in F606W$-$F814W \citep{vanDokkum_22b} and converted to the $g-i$ system applying a shift of 0.3 mag. FCC~224 has colors measured in $g_{475W}-i_{814W}$ \citep{Tang_25}, which are sufficiently close to $g-i$ for the purposes of this comparison.

\begin{figure}
\centering
\includegraphics[width=\columnwidth]{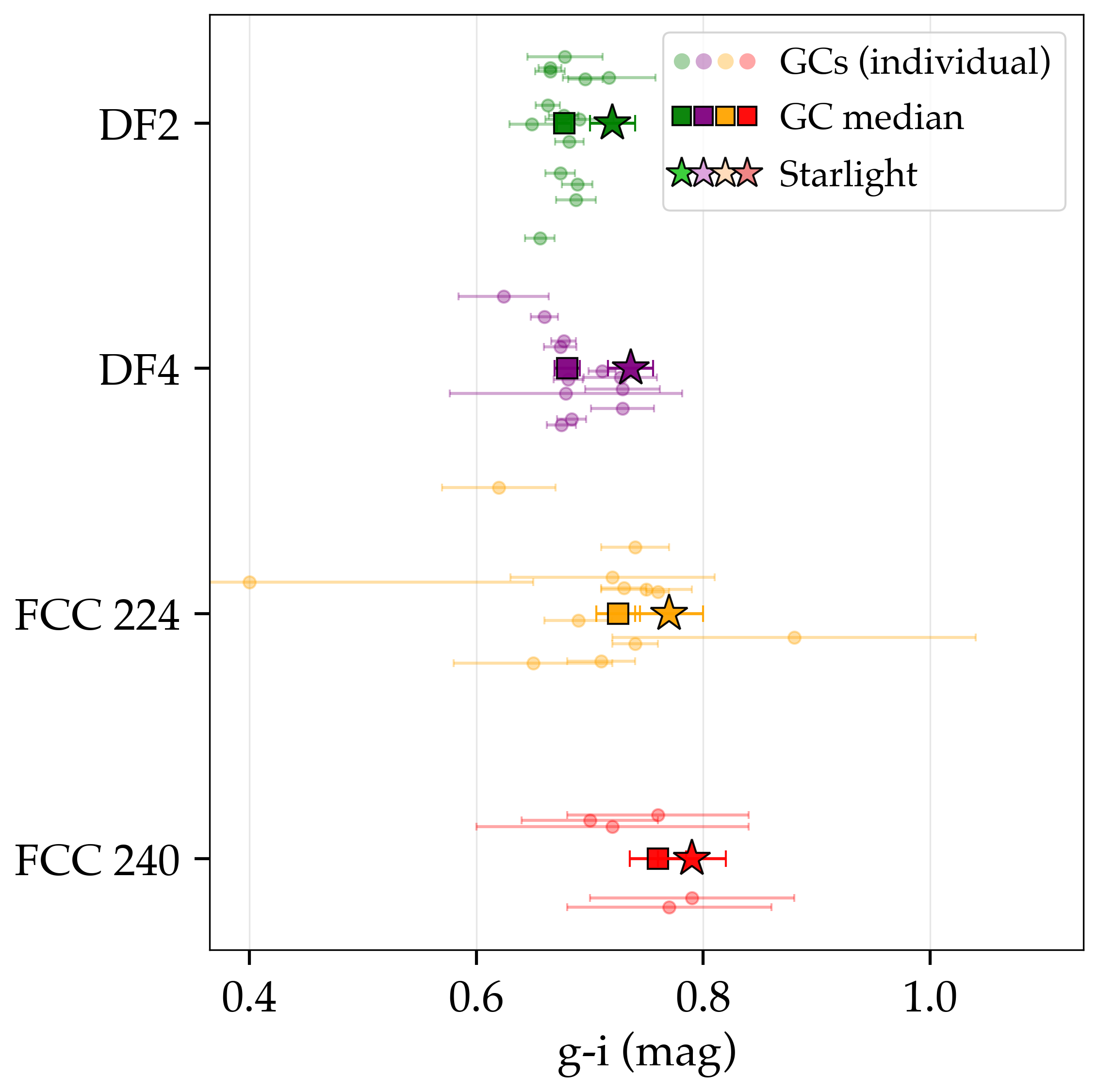}
\caption{
Color comparison between the galaxies and their GCs. 
Star markers show the integrated light of the galaxies, squares show the median GC colors, and small circles show the colors of the individual GCs, with small vertical offsets applied for visibility.
The GCs have a similar color to that of the stellar body within the uncertainties, although the galaxies are always slightly redder, indicating further chemical enrichment. 
The agreement between the stellar body and the GCs supports a common formation event.
}
\label{fig:stars_gcs_comparison}
\end{figure}

As shown in Figure~\ref{fig:stars_gcs_comparison}, the GCs in all four systems occupy a very narrow color range. As previously discussed by \citet{vanDokkum_22b} and \citet{Tang_25}, this tight distribution is particularly striking for DF2, DF4, and FCC~224. The same appears to be true for FCC~240, although the larger photometric uncertainties prevent a definitive conclusion. In all cases, the colors of the GCs are consistent with those of the diffuse stellar bodies within the uncertainties. The stellar bodies are systematically slightly redder than their GC systems, consistent with the findings of \citet{Fensch_19} that the GCs in DF2 were coeval but slightly more metal-poor than the stars.

This behaviour differs from that typically observed in dwarf galaxies and ultra-diffuse galaxies, where GC systems are usually more metal-poor and older than the bulk of the field stars, reflecting extended or multiple episodes of star formation, although this can vary depending on the environment. For example, \citealt{Janssens_24} found a color offset between the GCs and their host galaxies of just $0.07 \pm 0.08$ for a sample of dwarf galaxies in the Perseus cluster, consistent with the findings of \citealt{Janssens_22} for DGSAT~I and of \cite{Saifollahi_22} for a few Coma UDGs. In the present systems, the small color offset between the GCs and stellar bodies is more naturally interpreted as modest self-enrichment in the host galaxy following the initial star formation event. Stellar mass segregation could contribute to the color offset, but its expected impact on integrated galaxy colors is small compared to the effects of self-enrichment. In this picture, the GCs formed during the first, intense burst of star formation triggered by the collision, while star formation may have continued briefly at lower efficiency, enriching the stellar body without forming additional long-lived clusters.

The similarity between GCs and stars is further supported by the spectroscopic measurements presented in Section~\ref{sec:data}. Taken together, the agreement in colors, ages, and metallicities indicates that the stars and GCs in each galaxy formed during the same event. 

\section{Morphology and outer structure}
\label{sec:morphology}

\begin{figure*}[!ht]
\centering
\includegraphics[width=0.9\textwidth]{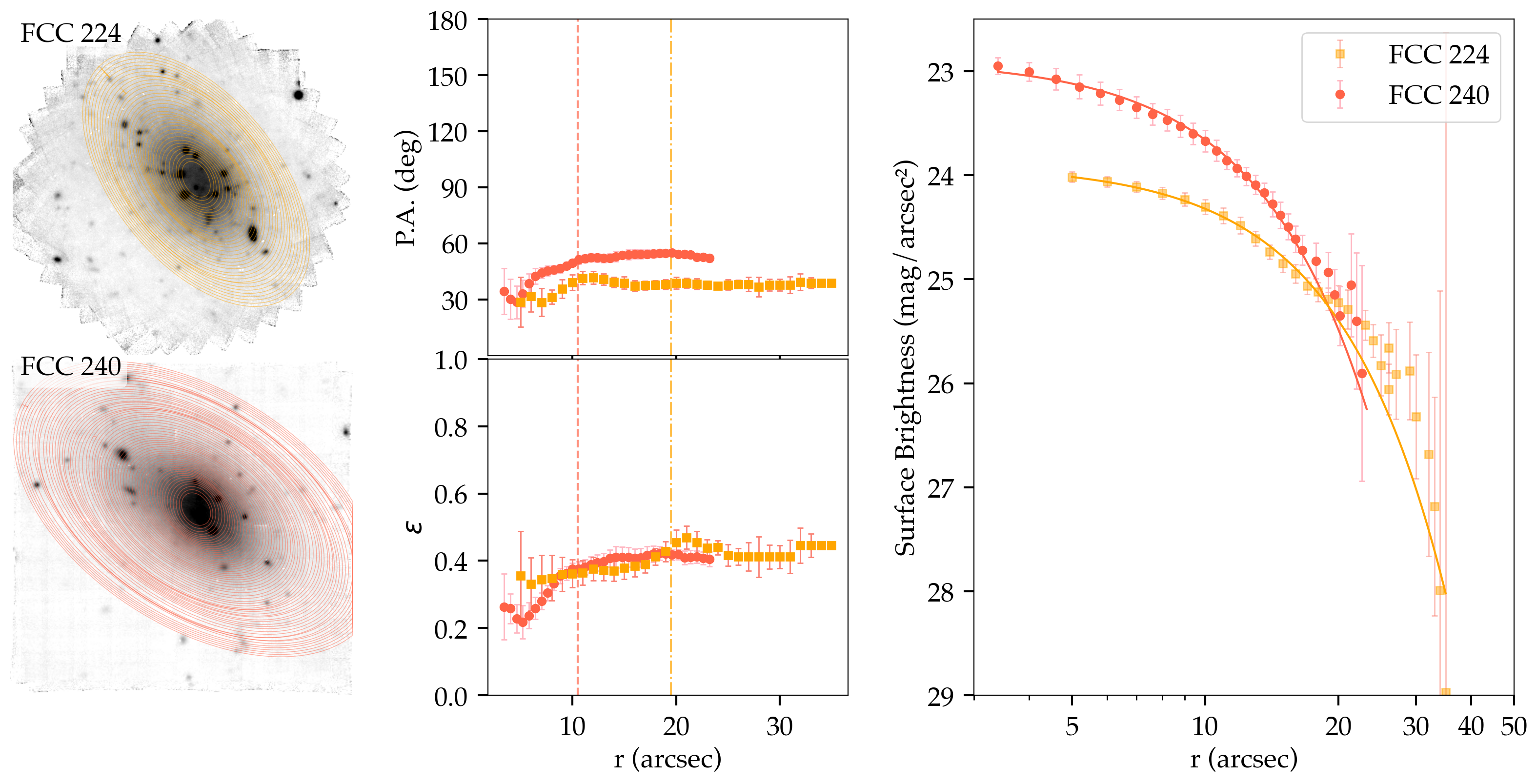}
\caption{
Isophotal analysis of FCC~224 and FCC~240. 
Left: MUSE pseudo-$g$-band images with elliptical isophotes. 
Middle: radial profiles of position angle and ellipticity. Both galaxies show stable shapes across most radii and remain closely aligned. The red dashed and orange dash-dotted lines show the $R_{\rm e}$ of FCC~240 and FCC~224, respectively.
Right: surface brightness profiles with best-fitting S\'ersic models. At large radii both galaxies begin to show an upward bending with relation to the inner S\'ersic profile, similar to DF2 and DF4 \citep{Keim_22}. These deviations may reflect weak tidal effects, projection effects in mildly non-axisymmetric systems, or a combination of both.}
\label{fig:isophotes}
\end{figure*}

We now examine the structure of FCC~224 and FCC~240 to search for tidal signatures and to test whether their morphologies are consistent with a shared origin. Following the method of \cite{Keim_22} and using \texttt{Ellipse} modelling in \texttt{photutils} \citep{photutils}, we used deep imaging to extract radial isophotes and to measure the ellipticity and position angle profiles of the galaxies (see Figure~\ref{fig:isophotes}).

For this analysis, we constructed pseudo--$g$-band images from the fully reduced MUSE datacubes by integrating the flux over the wavelength range corresponding to the DECaLS DR10 $g$-band transmission curve. Since these pseudo--$g$ images are not photometrically calibrated by default, we calibrated their zeropoints by matching the MUSE and DECaLS $g$-band profiles over the radial range where the galaxy light dominates, following \citet{Mirabile_25}. Background subtraction was applied to both images using the \texttt{MedianBackground} algorithm from \texttt{photutils}. The recovered zeropoint is 35.3 mag. GCs and other background/foreground sources were masked for this analysis, and the masked regions were filled with the median value computed from the neighbouring pixels.

The isophotes of both galaxies are smooth and concentric across most radii. There are no signs of bars, bulges, inner asymmetries, or embedded disks. Their ellipticity and position angle profiles are also very similar. The shapes of the galaxies remain nearly constant over the region where the light is well described by a S\'ersic profile, suggesting that both systems have undergone similar structural evolution. Both systems are relatively round in their inner regions, with ellipticities of $\sim0.2-0.3$, and become more flattened toward $R_{\rm e}$, consistent with the $\lambda_{R_{\rm e}}$ analysis presented in Appendix~\ref{sec:lambda_appendix}.

At large radii, the surface-brightness profiles begin to deviate from the best-fitting S\'ersic models and the outer isophotes show small departures from pure ellipses. The deviations appear as an upward bending, i.e. an excess of light at large radii. This behavior is qualitatively similar to what was observed in DF2 and DF4 by \cite{Keim_22}, where the excess light at faint surface-brightness levels was interpreted as possible signatures of tidal distortion. The onset of these deviations in FCC~224 and FCC~240 occurs at $\mu_g \simeq 25.5-26.5~\mathrm{mag\,arcsec^{-2}}$. At these surface-brightness levels, the outer stars are only weakly bound and are therefore more sensitive to tidal forces in galaxies with shallow gravitational potentials.

At the same time, we caution that these outer features are not unique evidence for ongoing tidal stripping. A weak position-angle twist or outer deviation from pure ellipses can also arise through projection in mildly non-axisymmetric or triaxial systems, especially when the surface-brightness gradient becomes shallow \citep{Mihalas_Binney_81}. In addition, this interpretation depends sensitively on the adopted sky background, as even small over- or under-estimates can produce apparent down- or upward bending in the outer light profiles. We therefore regard the outer isophotal structure as suggestive of disturbance, but not conclusive about tides.

Another clear feature is the alignment of the position angles of FCC~224 and FCC~240, varying by less than 15 degrees between the two galaxies. A similar situation is observed among the dwarf galaxies in the ``bullet trail'' in the NGC~1052 group, as reported by \cite{Tang_24}, where the position angles of the dwarfs are not perfectly parallel to the trail but show a clear tendency to be approximately aligned. In that study, this behaviour was interpreted as evidence for a shared origin or a common tidal history. The relative alignment seen in FCC~224 and FCC~240 therefore points toward a similar possibility, although alignment alone is not uniquely diagnostic of the underlying formation mechanism.

\section{Discussion: A Second Bullet-Dwarf Pair in the outskirts of the Fornax Cluster?}
\label{sec:discussion}

The discovery of a second pair of galaxies with unusually low inner dark matter content and overluminous GCs outside the NGC~1052 group suggests that the formation process behind DF2 and DF4 may not be unique. Across Sections \ref{sec:darkmatter}--\ref{sec:morphology}, we showed that FCC~224 and FCC~240 share the main internal properties of the NGC~1052 dwarfs, also summarized in Table~\ref{tab:comparison}: both are old ($\sim$10~Gyr), metal-poor ($\mathrm{[M/H]} \sim -1$), and have very low stellar and GC velocity dispersions. Their GCLFs are top-heavy, and their diffuse stellar light and GC systems have consistent stellar populations. Within one effective radius, their dynamical masses are consistent with the stellar masses, leaving little room for a dominant dark matter component in the inner regions. These properties are broadly consistent with the expectations from the bullet-dwarf scenario \citep{Silk_19}.

\begin{table*}[!ht]
\centering
\caption{Comparison of Bullet-Dwarf Candidate Pairs}
\begin{threeparttable}
\scalebox{0.9}{
\begin{tabular}{l|cc|cc}
\hline
Property & NGC~1052-DF2 & NGC~1052-DF4 & FCC~224 & FCC~240 \\
\hline
Distance (Mpc) & $22.1 \pm 1.2$ & $20.0 \pm 1.6$ & $18.6 \pm 2.7$ & $\sim20$ \\
$M_{\star}$ ($10^8 M_\odot$) & 2.0 & 1.5 & 1.7 & 0.8 \\
$R_{\rm e}$ (kpc) & 2.7 & $2.1 \pm 0.1$ & $1.89 \pm 0.01$ & $1.0 \pm 0.1$ \\
$\mu_{g,0}$ (mag arcsec$^{-2}$ ) & $25.08 \pm 0.02$ & $24.95 \pm 0.02$ & $24.69 \pm 0.03$ & $22.78 \pm 0.13$ \\
$n$ & $0.55 \pm 0.02$ & $0.79 \pm 0.01$ & $0.75 \pm 0.02$ & $0.80 \pm 0.01$ \\
$b/a$ & $0.89 \pm 0.01$ & $0.87 \pm 0.01$ & $0.64 \pm 0.01$ & $0.57 \pm 0.01$ \\
PA (deg) & $-43.13 \pm 0.54$ & $-82.50 \pm 0.68$ & $37.8 \pm 0.5$ & $51.0 \pm 0.4$ \\ 
Age (Gyr) & $8.9 \pm 1.5$ & $10.7^{+1.4}_{-1.5}$ & $10.1 \pm 1.1$ & $10.0 \pm 1.3$ \\
Stellar [M/H] (dex) & $-1.1 \pm 0.1$ & $-1.0^{+0.1}_{-0.1}$ & $-1.0 \pm 0.2$ & $-0.8 \pm 0.3$ \\
$\langle g-i\rangle_{\star}$ & $0.72 \pm 0.03$ & $0.74 \pm 0.04$ & $0.77 \pm 0.05$ & $0.79 \pm 0.03$ \\
$\langle g-i\rangle_{\rm GCs}$ & $0.68 \pm 0.01$ & $0.69 \pm 0.03$ & $0.70 \pm 0.10$ & $0.74 \pm 0.03$ \\
$\sigma_{\star}$ (km s$^{-1}$) & $8.5_{-3.1}^{+2.3}$ & $8.0^{+2.3}_{-1.9}$ & $11.8 \pm 7.8$ & $13.2 \pm 9.2$ \\
$\sigma_{\mathrm{GCs}}$ (km s$^{-1}$) & $7.8^{+5.2}_{-2.2}$ & $4.2^{+4.4}_{-2.2}$ & $8.5^{+3.9}_{-2.7}$ & $9.2^{+9.6}_{-5.6}$ \\
Projected Separation at 20 Mpc (kpc) & \multicolumn{2}{c|}{240} & \multicolumn{2}{c}{75} \\
$\Delta V$ (km s$^{-1}$) & \multicolumn{2}{c|}{358} & \multicolumn{2}{c}{16} \\
Observed Configuration & \multicolumn{2}{c|}{Extended trail ($>$2 Mpc)} & \multicolumn{2}{c}{Tight pair} \\
\hline
\end{tabular}}
\begin{tablenotes}
      \small
      \item \textbf{Note.} Distances come from \cite{Shen_21} for DF2 and DF4, and from \cite{Tang_25} for FCC~224. FCC~240 is assumed to be at the distance of the Fornax cluster. The stellar populations of DF4 come from \cite{Tang_24}, and for DF2 from \cite{Fensch_19}. 
\end{tablenotes}
\end{threeparttable}
\label{tab:comparison}
\end{table*}

At the same time, FCC~224 and FCC~240 differ significantly from DF2 and DF4 in several aspects. 
While the internal properties are similar, the overall structure of the systems is not: DF2 and DF4 are relatively round and widely separated, whereas FCC~224 and FCC~240 are more flattened and form a much closer pair (Table~\ref{tab:comparison}). This suggests that the DF2/DF4 pair may not be the most appropriate comparison.
The NGC~1052 trail also contains several other dwarfs with similar stellar populations and closer projected separations \citep{Buzzo_23, Gannon_23, Tang_24, vanDokkum_22b, Keim_26}, which may offer more relevant comparisons. In particular, DF9 lies very close to DF4, with a separation comparable to that between FCC~224 and FCC~240.

DF9 has now also been shown to share the same low velocity dispersion as DF2 and DF4. Using Keck/KCWI, \citet{Keim_26} measured $\sigma_\star = 6.4^{+4.0}_{-4.3}$~km~s$^{-1}$, consistent with the velocity dispersion expected from its stellar mass alone and far below the expected for a standard dark matter halo. They therefore identified DF9 as a third dark-matter-deficient galaxy associated with the NGC~1052 trail. This result strengthens the broader picture that the trail contains multiple galaxies with unusually low inner dark matter content, and it makes the DF4/DF9 pair a particularly relevant comparison for the FCC~224/FCC~240 system.

Unlike DF2 and DF4, however, DF9 does not show as obvious an abundance of overluminous GCs \citep{Buzzo_23}. Instead, it hosts a compact central star cluster which resembles the GCs around DF2 and DF4 in luminosity, size, and overall stellar population properties \citep{Gannon_23,Keim_26}. This contrast does not necessarily require a different origin: it may indicate that similar formation pathways can leave behind different GC configurations, with some galaxies retaining prominent bright GC systems and others ending up with a more centrally concentrated star cluster component. Importantly, DF9, like FCC~240, is more compact than the UDG threshold and therefore provides a particularly useful comparison for the Fornax pair. Placing FCC~224 and FCC~240 in this broader context, the Fornax pair may likewise represent a different stage or outcome of a similar process, where the internal stellar populations and dynamics remain similar but the spatial arrangement and structural properties vary from system to system.

From this perspective, FCC~224 and FCC~240 may represent only a small portion of a more extended structure, analogous to a localized segment of the NGC~1052 trail, raising the possibility that additional related dwarfs remain undiscovered at larger separations. We can thus trace a line between the two galaxies, similarly to what was done by \cite{vanDokkum_22} to define a preferential direction to search for additional members. For this analysis, we assume that both galaxies are at the distance of the Fornax cluster (20~Mpc), as the SBF distance recovered by \cite{Tang_25} is formally consistent with 20~Mpc within the uncertainties. The projected positions of FCC~224 (RA = $54.8870^\circ$, Dec = $-31.7915^\circ$) and FCC~240 (RA = $55.1138^\circ$, Dec = $-31.6861^\circ$) define a straight line on the sky, which can be expressed as:
\begin{equation}
\Delta\mathrm{Dec} = 0.4647\Delta\mathrm{RA},
\end{equation}
where $\Delta\mathrm{RA} = \alpha - \alpha_{\rm FCC224}$ and $\Delta\mathrm{DEC} = \delta - \delta_{\rm FCC224}$. We use this line as an approximation of the projected trajectory of the system, while noting that the true trajectory is not known and could deviate from a straight line, for example if the interaction was not head-on.
If we assume, based on the age of FCC~224 and FCC~240, that the collision took place about 10~Gyr ago with a speed of about $300~\mathrm{km \, s^{-1}}$ \citep[the minimum required to separate baryons and dark matter; see e.g.][]{Lee_24}, the dark matter halos would now lie roughly 3~Mpc away along this axis. At the distance of Fornax, and assuming that the motion is mostly in the plane of the sky (as suggested by the small velocity separation between FCC~224 and FCC~240), this corresponds to about $\pm9$ degrees. This region is therefore the natural place to search for any dark-matter-dominated remnants or additional fragments.

At present, FCC~240 was identified primarily because of its close proximity to FCC~224, rather than as part of a systematic search for a larger population of related objects. It is therefore possible that additional low-surface-brightness dwarfs lie farther along the same projected trajectory but have not yet been identified or studied in detail. In the NGC~1052 system, multiple dwarfs were only recognized as part of a coherent structure after deep imaging over a wide area. A similar situation may apply in Fornax, where a more complete census of faint galaxies along the projected path could reveal further fragments of the same event. For example, FCC~122 lies close to the galaxies in projection, being 1.4 degrees away from FCC~224 ($\sim 0.5$ Mpc at 20 Mpc), and close to the projected axis (only $\sim 14$ kpc away) and may warrant closer inspection in future work.

Importantly, the environment of the FCC~224/FCC~240 system may also play an important role. Both galaxies lie on the outskirts of the Fornax cluster, in a region with a somewhat similar density to that of a group like NGC~1052. This lower density environment reduces the likelihood of strong tidal encounters after the collision, allowing these systems with shallow gravitational potentials to survive for several Gyr.

\begin{figure}
\centering
\includegraphics[width=\linewidth]{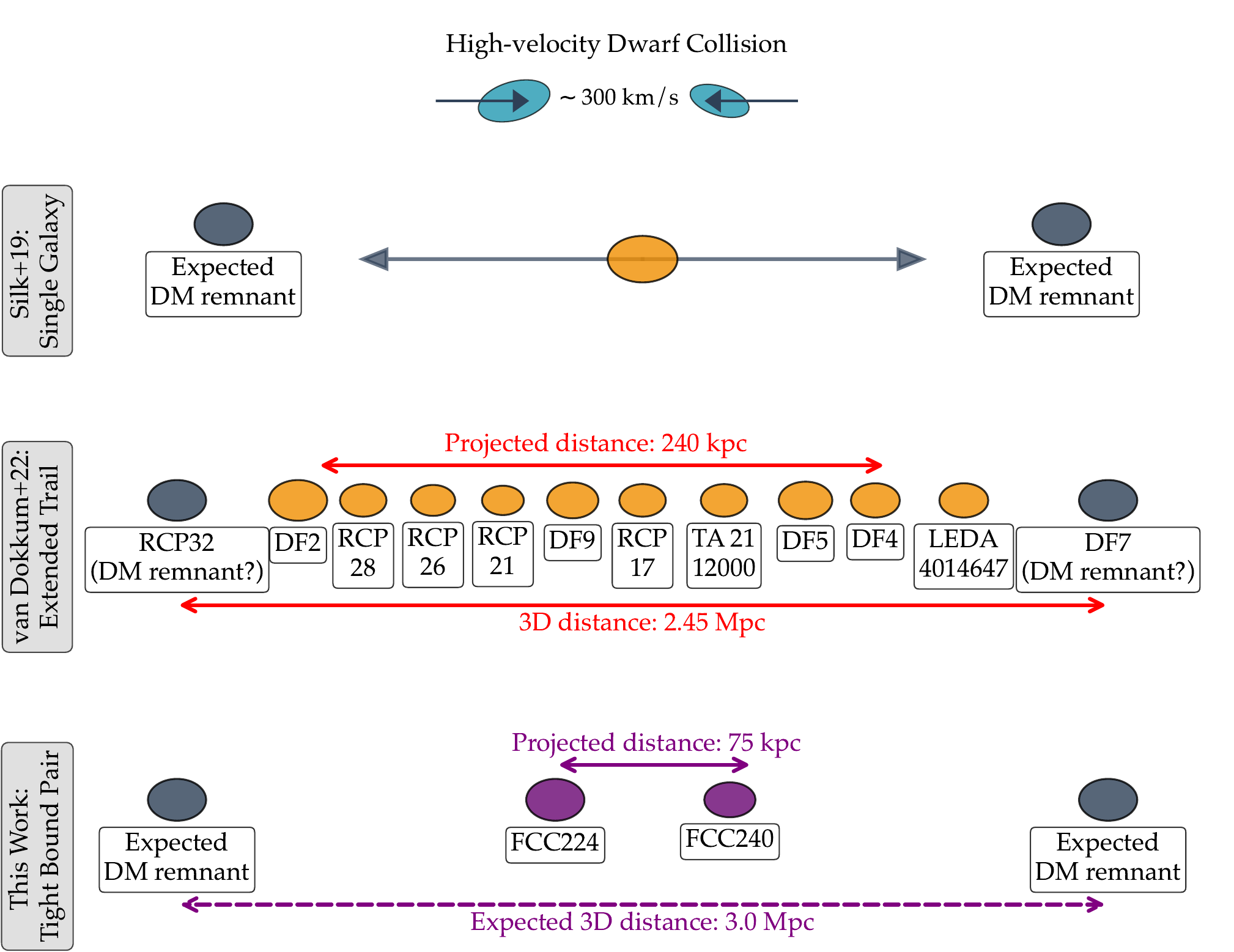} 
\caption{
Schematic illustration of possible outcomes of the bullet dwarf collision. 
\emph{Top:} Concept of a $\sim300~\mathrm{km\,s^{-1}}$ collision from \cite{Silk_19}. The baryonic matter forms a galaxy while the collisionless dark matter lags behind, creating a spatial offset. 
\emph{Middle:} The extended sequence of fragments in the NGC~1052 system, including DF2 and DF4, proposed by \citealt{vanDokkum_22} (only the original galaxies in \cite{vanDokkum_22} are included). The projected distance between DF2 and DF4 is about 240 kpc and the expected dark matter remnant lies at the edges of the trail. 
\emph{Bottom:} The FCC~224 and FCC~240 system. Here the initial conditions, assuming that the galaxies were indeed formed in a bullet-dwarf collision, end in a compact and long-lived bound pair separated by 75 kpc. The expected dark matter remnants are indicated. 
}
\label{fig:collision_scenarios}
\end{figure}

From a chronological perspective, \cite{Silk_19} first proposed that high-speed dwarf–dwarf collisions (miniature analogues of the Bullet Cluster) could separate baryons from dark matter, producing a dark-matter-deficient stellar remnant and creating the high-pressure conditions required to form unusually massive clusters. This idea was subsequently explored in detail by hydrodynamical simulations, which demonstrated that such collisions can generate multiple dark-matter-deficient fragments along a common trajectory rather than a single remnant \citep{Shin_20}. Motivated by these predictions, deep imaging of the NGC~1052 group revealed an extended, near-linear trail of dwarf galaxies spanning more than 2~Mpc, with DF2 and DF4 as two prominent members \citep{vanDokkum_22}. We now identify FCC~224 and FCC~240 in a very different configuration from the NGC~1052 system, yet exhibiting remarkably similar internal properties. This suggests that either both systems are explained by another scenario altogether that we yet do not know, or that the same basic mechanism of the bullet-dwarf may also manifest in different present-day configurations, i.e., instead of a single remnant, or an extended chain of fragments, it may leave behind a compact and long-lived bound pair. A third, more conservative interpretation is that FCC~224 and FCC~240 form a chance pair of dwarf galaxies with similar properties but unrelated formation histories, a possibility that can only be tested with deeper and more comprehensive data.

This sequence of theoretical works and hypotheses have been closely followed by hydrodynamical simulations \citep{Shin_20,Lee_21,Lee_24, Wang_25}, which reinforce that the possible outcomes of this collision are not unique: relatively small variations in the geometry and orbital energy of the encounter can yield either a stretched sequence of multiple fragments or a system where only a few massive remnants survive as a bound pair. For example, in the trail-like case observed in the NGC~1052 group, the simulations by \cite{Lee_24} show that an initially bead-like sequence can be further elongated over time by the gravitational influence of the separating progenitors at its ends, while projection effects can amplify the apparent contrast between different systems. In this context, FCC~224 and FCC~240 may represent one of several possible observational configurations produced by the same underlying process, and Figure~\ref{fig:collision_scenarios} provides a schematic summary of this progression from the original collision picture to the diversity of outcomes now suggested by both observations and simulations.

The discovery of FCC~224 and FCC~240 opens the possibility of identifying more galaxies of this type. A natural next step is to search for additional stellar fragments or dark matter-dominated remnants along the projected trajectory defined by the pair. Searches for more cases in upcoming wide-area surveys such as \emph{Euclid} will also be important, particularly as these data will improve the census of low-surface-brightness dwarfs and their GC systems. Obtaining full star formation histories will also help trace the evolutionary history of these galaxies. On the theoretical side, hydrodynamical simulations with higher resolution and a broader exploration of initial conditions will be important, as will quantifying how often dwarf--dwarf collisions with relative velocities $\gtrsim300~\mathrm{km \, s^{-1}}$ occur.

FCC~224 and FCC~240 therefore strengthen the case that high-velocity collisions can form galaxies with low inner dark matter content and overluminous GC populations, while also showing that the present-day configuration of the remnants can vary substantially. Other processes such as tidal stripping \citep{Ogiya_18,Ogiya_22, Moreno_22}, or feedback-driven expansion \citep{Trujillo_Gomez_21} may play a role in shaping these galaxies, but none of them naturally explain both the low inner velocity dispersions and the bright GC systems observed here and in DF2 and DF4. The Fornax pair therefore moves the discussion from whether such galaxies exist to how common these events are, and how their observational properties vary across the dwarf galaxy population.

\section{Conclusions}
\label{sec:conclusion}

We have presented deep VLT/MUSE spectroscopy of the ultra-diffuse galaxy FCC~224 and its close companion FCC~240 in the outskirts of the Fornax cluster. The results from this work show that they may form a second example of a pair of galaxies with unusually low inner dark matter content and overluminous GCs, and are therefore possible analogues of NGC~1052--DF2 and DF4. Throughout the paper we examined the internal kinematics, GC systems, stellar populations, and structural properties of both galaxies, and all of these measurements point to a close match between the Fornax and NGC~1052 systems. In this case, their association with the Fornax cluster leaves less room for distance-related controversies than in the NGC~1052 group, where such issues have been extensively debated.

The velocity dispersions measured from the stars and GCs indicate that both galaxies lack a dominant dark matter halo within one $R_{\rm e}$, mirroring what is observed in DF2 and DF4. Their dynamical masses are consistent with the stellar masses and fall below the values expected for normal dwarf galaxies hosted by typical cuspy or cored dark matter halos. While their significant flattening and lack of rotation suggest that anisotropy and possibly triaxiality are important, these effects are not expected to change the mass within the half-light radius enough to alter this conclusion. The GC systems of both galaxies are also unusual, showing an excess of overluminous GCs which closely resemble those in DF2 and DF4.

The colors and stellar population properties of the diffuse light and the GCs are also consistent with a single formation event. They are old, with ages of about 10~Gyr, and have metallicities around $\mathrm{[M/H]}\sim -1.0$ dex. The close agreement between the stellar and GC populations supports a common origin, with the stellar body being slightly redder likely due to self-enrichment. 

The structural analysis of FCC~224 and FCC~240 shows that they share similar shapes and orientations. Their isophotal profiles follow each other closely and show mild upward bendings at large radii, at surface-brightness levels comparable to those where DF2 and DF4 begin to deviate from smooth S\'ersic profiles. These outer features may reflect weak tidal effects, projection effects in mildly non-axisymmetric systems, or a combination of both. The alignment of their position angles further supports a shared history. Together with their small projected separation of 75~kpc and low relative velocity of 16~km~s$^{-1}$, FCC~224 and FCC~240 form a compact pair that has likely remained bound for most of its lifetime.

Taken together, these results suggest that FCC~224 and FCC~240 may have formed in a high-velocity collision similar to the event proposed for DF2 and DF4. The existence of this second pair hints at the possibility that the process is repeatable and not unique to one group, and suggests that the range of possible outcomes is broad: a collision may produce an extended sequence of fragments, or it may leave behind a compact, long-lived bound pair.

The discovery of FCC~224 and FCC~240 opens the possibility of identifying more galaxies of this type. A natural next step is to search for additional stellar fragments or dark-matter-dominated remnants along the trajectory defined by the pair. On the theoretical side, simulations with higher resolution and a broader exploration of initial conditions will be important. Together, these new observations and models will help establish how common these events are, and provide a clearer view of how this unusual class of galaxies forms.

\section*{Acknowledgments}
We are thankful to the anonymous referee for their insightful suggestions which greatly improved the manuscript. We are also thankful to Eric Emsellem and Lodovico Coccato for the help with reducing and combining the deep MUSE cubes. DF thanks the ARC for financial support via DP250101673. AJR was supported by National Science Foundation grant AST-2308390.

\bibliography{bibli_dedup}{}
\bibliographystyle{aasjournalv7}

\appendix

\section{The main properties of FCC~240}
\label{sec:appendix_fcc240}

In this Appendix we describe the procedure used to extract the structural parameters of FCC~240, to estimate its stellar mass, and to identify its GC candidates from the DECaLS imaging. All steps follow the same methodology adopted in our previous works for FCC~224 and ultra-diffuse galaxies \citep[][]{Tang_25, Buzzo_24,Buzzo_25a}. We also adopt the standard GC selection criteria used in \citet[][their Section 3]{Buzzo_23}.

We used the publicly available DECaLS DR10 $g$, $r$, and $i$ bands. FCC~240 lies at the boundary between two DECaLS fields, which makes the analysis more challenging. Because of this, the photometric modelling was carried out on the field that covers the largest fraction of the galaxy. The GC candidate selection was performed in both fields independently, and duplicates were removed. This approach avoids the significant depth variation that appears in the combined image, where the overlap region becomes deeper than the outer parts. The photometry for the galaxy and GC candidates has been corrected for Galactic extinction using \citet{Schlafly_11}.

\subsection{Structural and photometric properties}

The galaxy modelling was performed with \texttt{GALFITM} \citep{Haussler_13} in the same way as in earlier work on ultra diffuse galaxies \citep[e.g.,][]{Buzzo_24,Buzzo_25a}. All foreground and background sources were masked during the fit, including the GC candidates. We adopted a single Sérsic model where the magnitudes were allowed to vary independently in each band, while the effective radius, Sérsic index, axis ratio, and position angle were kept fixed across all three. The construction of the point spread function, the background characterization, and the masking strategy follow \citet{Buzzo_24}. 

\begin{figure}[!h]
\centering
\includegraphics[width=0.76\linewidth]{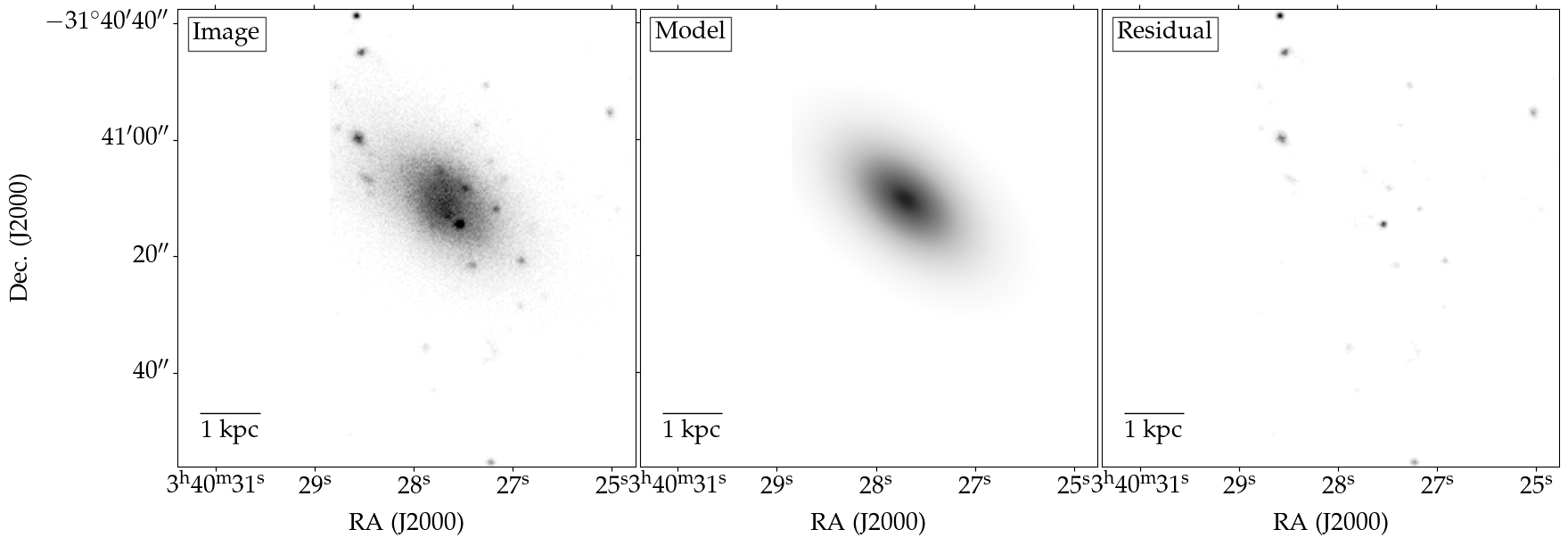}
\caption{
GALFIT modelling of FCC~240 in the $g$ band. From left to right: original DECaLS image, best fitting Sérsic model, and residual map.
}
\label{fig:galfit_fcc240}
\end{figure}

The resulting \texttt{GALFITM} model for FCC~240 is shown in Figure~\ref{fig:galfit_fcc240}. The structural parameters are consistent with the isophotal fitting analysis in Section~\ref{sec:morphology}. The final values are listed in Table~\ref{tab:FCC240}. All magnitudes are given in the AB system.

After obtaining the integrated magnitudes in all bands, we estimated the stellar mass of FCC~240 using \texttt{Prospector}, with the same configuration used in \citet{Buzzo_24} and \citet{Buzzo_25a}. The best fitting model gives a stellar mass of $M_{\star} = 8.1 \times 10^{7}\,M_\odot$. This value is used throughout the paper. The stellar populations obtained with \texttt{Prospector} are consistent within the uncertainties with the results obtained with spectroscopy and used throughout the paper.

\subsection{GC Candidate Selection}

Point sources around FCC~240 were detected with \texttt{SExtractor} in dual image mode, adopting the composite $g+r+i$ frame as the detection image. Sources with at least five connected pixels above a $2\sigma$ threshold were included.  

Following \citet{Buzzo_23}, we applied a combination of morphological and photometric criteria to identify likely GCs. At the distance of the Fornax cluster, GCs are unresolved in the DECaLS images. We therefore selected sources with ellipticity $\epsilon < 0.3$, magnitudes in the range $-11.2 < M_i < -5.2$ mag (i.e., three sigma from the GCLF peak), and colors consistent with simple stellar population models of old ($\sim 10$ Gyr), metal poor clusters in $g-r$ and $g-i$. Only sources within three effective radii of the galaxy were considered. To estimate the contamination, we applied the same criteria to background regions within the field. We expect two contaminants within the area corresponding to three effective radii of the galaxy.

Applying these criteria, we identified six GC candidates, five of which have had their membership confirmed with MUSE in Section \ref{sec:data}, and the sixth is likely a background galaxy given its redder color and relative velocity with respect to FCC~240. The properties of all six are summarized in Table~\ref{tab:FCC240}.

We estimated the completeness by injecting 10000 artificial stars into the images and repeating the detection and selection process. We find that our analysis is 90\% complete down to $i = 24.0$ mag. We use this completeness to estimate the total GC population of FCC~240, assuming a standard Gaussian GCLF \citep{Jordan_15}. This provides a conservative upper limit on the GC population, as we avoid adopting a top-heavy GCLF whose shape depends on the assumed distance. We estimate $N_{\rm GC}\simeq8\pm3$, where the uncertainty comes from Poisson statistics.

Following recent studies of dwarf galaxy GC systems using MUSE \citep[e.g.,][]{Mirabile_25}, we also explored the possibility of estimating the GC population directly from the pseudo-$g$, $r$, and $i$ images reconstructed from the datacube. However, FCC~240 occupies a large fraction of the MUSE FoV, making background estimation and completeness corrections substantially more uncertain than in \cite{Mirabile_25}. In addition, the MUSE observations for FCC~240 reach a $g$-band $5\sigma$ depth of 23.3 mag, comparable to DECaLS DR10 \citep[23.95 mag;][]{Inchausti_25}. Our tests with the MUSE images therefore did not improve the constraints on the faint end of the GCLF. Nonetheless, we expect a population of fainter GCs to exist, and deeper imaging will be necessary to properly characterize it.

\begin{table}[!h]
    \centering
    \caption{Structural and photometric properties of FCC~240 and its confirmed GCs}
    
    \begin{tabular}{c|c|cccccc}
        \textbf{Property} & \textbf{FCC~240} & \textbf{BG1} & \textbf{GC2} & \textbf{GC3} & \textbf{GC4} & \textbf{GC5} & \textbf{GC6} \\ \hline
        RA (J2000) & 55.1138 & 55.1148 & 55.1133 & 55.1136 & 55.1127 & 55.1144 & 55.1116 \\
        Dec. (J2000) & $-31.6861$ & $-31.6842$ & $-31.6852$ & $-31.6869$ & $-31.6821$ & $-31.6865$ & $-31.6861$\\
        $g_0$ (mag) & $16.4 \pm 0.1$ & $22.4 \pm 0.1$ & $24.2 \pm 0.1$ & $23.2 \pm 0.1$ & $23.3 \pm 0.2$ & $24.6 \pm 0.1$ & $24.0 \pm 0.1$ \\
        $(g-i)_{0}$ (mag) & $0.79 \pm 0.08$ & $1.32 \pm 0.08$& $0.77 \pm 0.09$ & $0.70 \pm 0.06$ & $0.76 \pm 0.08$ & $0.72 \pm 0.12$ & $0.79 \pm 0.09$ \\
        S/N (\AA$^{-1}$) & 84 & 9.8 & 9.0 & 12.2 & 6.1 & 9.6 & 7.6 \\
        $V$ (km s$^{-1}$) & $1408.5 \pm 2.0$ & $1653 \pm 252$ & $1406 \pm 3$ & $1404 \pm 2$ & $1458 \pm 15$ & $1412 \pm 3$ & $1423 \pm 5$ \\ \hline
    \end{tabular}
    \label{tab:FCC240}
\end{table}

\section{GCs around FCC~224}

In this section, we list the recovered properties for the probed GCs around FCC~224.

\begin{table}[!h]
\centering
\caption{Spectroscopic properties of probed GC candidates around FCC~224 \citep[numbers follow][]{Tang_25}}
\label{tab:FCC224}
\begin{threeparttable}
\footnotesize
\begin{tabular}{c|cc}
\textbf{ID} & \textbf{S/N (\AA$^{-1}$)} & \textbf{$V$ (km s$^{-1}$)} \\ \hline
GCC1  & 2.3 & -- \\
GC2  & 5.4 & $1449 \pm 9$ \\
GC3  & 7.4 & $1412 \pm 4$ \\
GC4  & 7.7 & $1417 \pm 3$ \\
GC5  & 6.7 & $1429 \pm 7$ \\
GCC6  & 2.2 & -- \\
GC7  & 7.2 & $1425 \pm 3$ \\
GC8  & 6.7 & $1429 \pm 3$ \\
GCC9  & 3.3 & -- \\
GC10 & 6.9 & $1436 \pm 11$ \\
GCC11 & 5.5 & $1309 \pm 125$ \\
GC12 & 4.9 & $1423 \pm 11$ \\ \hline
\end{tabular}
\begin{tablenotes}
      \small
      \item \textbf{Note.} IDs starting with `GCC' are candidates, and with `GC' are confirmed.
\end{tablenotes}
\end{threeparttable}
\end{table}

\section{Angular momentum measurements}
\label{sec:lambda_appendix}

As discussed in the main text, FCC~224 and FCC~240 are flattened systems with no measurable rotation. In such cases, it is useful to quantify their projected angular momentum in order to assess whether their flattening could be supported by rotation or instead requires anisotropy or a more complex intrinsic structure. We therefore compute the angular momentum parameter $\lambda_{R_{\rm e}}$ for both galaxies, and compare them with DF2 and DF4.

We computed the projected angular momentum parameter $\lambda_{R_{\rm e}}$ following \citet{Emsellem_11}, summing over Voronoi bins within an elliptical aperture of semi-major axis $R_{\rm e}$. For comparison, we applied the same procedure to archival MUSE data for NGC~1052-DF2 and DF4. The ellipticity $\epsilon_{\rm e}$ is defined as a luminosity-weighted quantity within the same aperture, following \citet{Cappellari_07}.

A key limitation is that the velocity dispersions in individual Voronoi bins are not reliable for these galaxies. The integrated dispersions measured within $1R_{\rm e}$ are already well below the instrumental resolution of MUSE, and the bin-by-bin values are therefore dominated by noise and uncertainties in the LSF. For this reason, we do not use the individual dispersions from the Voronoi bins. Instead, we adopt the integrated velocity dispersion measured within $1R_{\rm e}$ for each galaxy and assign this value to all bins. The uncertainty on $\lambda_{R_{\rm e}}$ is then estimated using the upper and lower limits of the integrated dispersion. This results in large uncertainties, but avoids over-interpreting noisy measurements.

The moment-based ellipticity $\epsilon_{\rm e}$ differs from the isophotal ellipticity at exactly $R_{\rm e}$. Because it is luminosity-weighted, it is more sensitive to the inner regions of the galaxies. As shown in Fig.~\ref{fig:isophotes}, FCC~224 and FCC~240 are rounder in their inner parts, with ellipticities of about $0.2$--$0.3$, and become more flattened toward $R_{\rm e}$. This explains why the $\epsilon_{\rm e}$ values used in this diagram are smaller than those derived from the outer isophotes or from single S\'ersic fits.

\begin{figure}[h!]
\centering
\includegraphics[width=0.5\columnwidth]{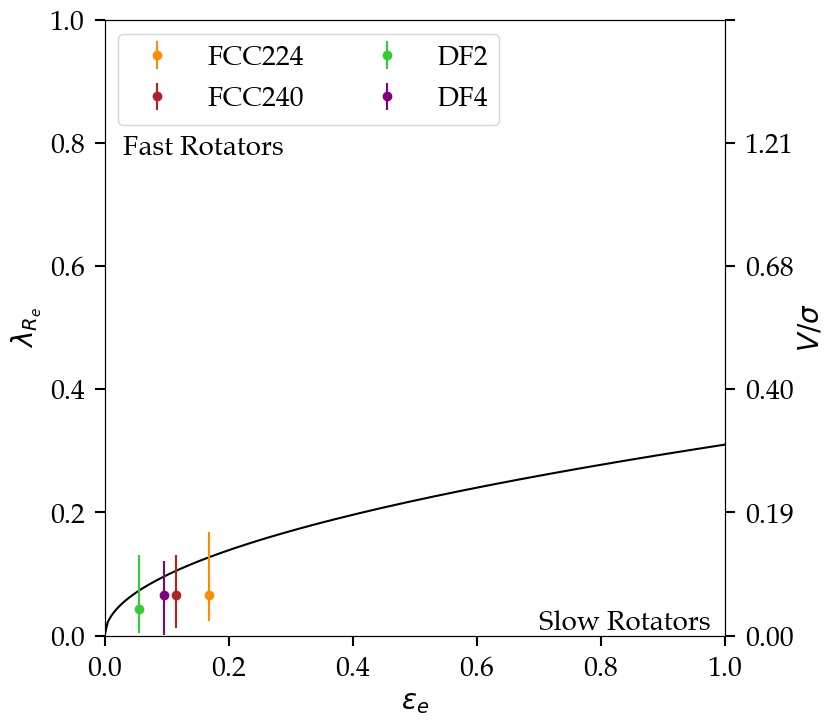}
\caption{Location of FCC~224, FCC~240, DF2, and DF4 in the $\lambda_{R_{\rm e}}-\epsilon_{\rm e}$ plane. The solid line marks the empirical fast--slow rotator division from \citet{Emsellem_11}, $\lambda_{R_{\rm e}}=0.31\sqrt{\epsilon_{\rm e}}$. The right axis shows the corresponding scale in $V/\sigma$ for comparison. Because the dispersions are below the MUSE instrumental resolution, the plotted uncertainties are dominated by the uncertainty on the integrated velocity dispersion.}
\label{fig:lambda_epsilon}
\end{figure}

The resulting $\lambda_{R_{\rm e}}-\epsilon_{\rm e}$ diagram for FCC~224, FCC~240, DF2, and DF4 is shown in Fig.~\ref{fig:lambda_epsilon}.

As shown in Fig.~\ref{fig:lambda_epsilon}, all four galaxies have low $\lambda_{R_{\rm e}}$ values, with $\lambda_{R_{\rm e}}<0.1$, consistent with the absence of measurable rotation in their velocity fields. Some individual error bars extend toward the fast-rotator boundary because the velocity dispersion uncertainties are large. We therefore use this diagram only as a consistency check, rather than as a precise measurement of intrinsic shape. The main result is that none of the galaxies shows evidence for rotation strong enough to explain its observed flattening.

\end{document}